\documentclass{egpubl}
 
\JournalSubmission    
\usepackage[T1]{fontenc}
\usepackage{dfadobe}  

\usepackage{cite}  
\usepackage{todonotes}
\BibtexOrBiblatex
\PrintedOrElectronic

\usepackage{amsmath}
\usepackage{amssymb}
\usepackage{stfloats}
\graphicspath{{images/SVG/}}
\usepackage{svg}
\usepackage{algorithm}
\usepackage{algpseudocode}

\definecolor{redNode}{rgb}{0.9, 0.3, 0.1}
\definecolor{blueNode}{rgb}{0.0, 0.5, 0.8}
\usepackage{fontawesome5}

\usepackage[bb=dsserif]{mathalpha}
\usepackage{bm}

\usepackage{subcaption}

\usepackage{tikz}
\usetikzlibrary{shapes, arrows, calc, arrows.meta, fit, positioning}

\tikzset{
	-Latex,auto,node distance = 1 cm and 1 cm, thick,
	centernode/.style ={circle, draw, fill=yellow, fill opacity=1, minimum width = 0.2 cm}, 
	nodePlus/.style ={circle, fill=redNode, fill opacity=0.8, minimum width = 0.2 cm},
	nodeMoins/.style ={circle, fill=blueNode, fill opacity=0.8, minimum width = 0.2 cm},
	point/.style = {circle, draw, inner sep=0.18cm, fill, node contents={}},
}

\title[Reeb Graph-based Mesh Segmentation]{Flexible Mesh Segmentation via Reeb Graph Representation of Geometrical and Topological Features}

\author[F. Beguet \& S. Lanquetin \& R. Raffin]
{\parbox{\textwidth}{\centering F. Beguet
,
        S. Lanquetin 
        and R. Raffin 
        }
        \\
{\parbox{\textwidth}{\centering University of Burgundy, LIB, EA7534\\
       }
}
}

\begin{document}


\maketitle
\begin{abstract}
This paper presents a new mesh segmentation method that integrates geometrical and topological features through a flexible Reeb graph representation. The algorithm consists of three phases: construction of the Reeb graph using the improved topological skeleton approach, topological simplification of the graph by cancelling critical points while preserving essential features, and generation of contiguous segments via an adaptive region-growth process that takes geometric and topological criteria into account. Operating with a computational complexity of $O(n \log n)$ for a mesh of $n$ vertices, the method demonstrates both efficiency and scalability. An evaluation through case studies, including part-based decomposition with Shape Diameter Function and terrain analysis with Shape Index, validates the effectiveness of the method in completely different applications. The results establish this approach as a robust framework for advanced geometric analysis of meshes, connecting the geometric and topological features of shapes.

  

\end{abstract}  
\section{Introduction}

Mesh segmentation is a fundamental process in computer graphics and geometric analysis. It decomposes 3D models into meaningful regions, enabling applications like CAD model manufacturing \cite{XIAO2011685, 17976}, shape matching \cite{lin2020seg}, topology optimization \cite{9525504}, texture mapping \cite{lai2018high, li2020novel}, and character animation \cite{yuan2016space, ma2024easyskinning}. While conceptually straightforward, effective implementation faces several fundamental challenges. The geometric complexity of shapes requires precise analysis of regions of high curvature, which can be affected by mesh irregularities and acquisition noise. The inherent diversity of geometric shapes further complicates the task: organic objects exhibit smooth transitions, while mechanical objects present distinct boundaries, necessitating adaptive approaches. Finally, a major difficulty consists in balancing over-segmentation (excessive partitioning) and under-segmentation (insufficient separation) while simultaneously processing local geometric and global topological features. 

\subsection{Related works}

This article covers two fundamentally different classes of approaches regarding mesh segmentation. Surface-based methods exploit local geometric properties to partition meshes into regions sharing similar features such as convexity or curvature, while skeleton-based methods extract a compact topological representation to guide the segmentation process. These classes of approaches have been widely studied \cite{attene2006mesh, agathos20073d, shamir2008survey, theologou2015comprehensive, rodrigues2018part}, each offering distinct advantages for different applications.

\subsubsection{Surface-based segmentation}

Surface-based mesh segmentation algorithms use geometric features like curvature or saliency to divide meshes into various parts. These methods are classified into three categories: region-growing, clustering, and boundary segmentation.

\textbf{Region-growing} methods \cite{zuckerberger2002polyhedral, lavoue2008markov, bergamasco2012graph, 9151831}, inspired by polyhedral surface decomposition \cite{chazelle1995strategies}, progressively enlarge a set of seeds on the mesh using predetermined criteria to form regions that align with needed geometrical features. Other approaches based on image analysis \cite{koschan2003perception, benjamin2011heat, lou2020watershed, xue2021watershed} partition meshes into \textit{watersheds} to improve region quality. The effectiveness of these methods is highly dependent on seeds selection and stopping criteria, as choices can lead to over- or under-segmentation, often requiring additional refinement steps.

\textbf{Clustering} methods analyze elements in a feature space defined by geometric and topological properties. tandard clustering \cite{shlafman2002metamorphosis, lai2008fast, shapira2008consistent, 17976} follows the principles of K-means \cite{lloyd1982least} by grouping features around representative centers. Some techniques incorporate spectral analysis \cite{liu2004segmentation, fang2011heat, zhang2012variational} to better determine optimal segment boundaries and their number. \textbf{Hierarchical clustering} proceeds either top-down \cite{katz2003hierarchical, liu2007mesh, zhang2015shape} by recursively subdividing the mesh according to size or geometric complexity, or bottom-up \cite{attene2006hierarchical, XIAO2011685, s23010416} by iteratively merging adjacent regions based on similarity metrics. Both hierarchical variants have difficulties in determining optimal termination criteria and computational efficiency for large-scale meshes.

\textbf{Boundary segmentation} methods identify segment contours through connected edges and vertices, following Hoffman's theory \cite{hoffman1997salience} of human visual perception of object boundaries. Approaches include shortest path computation \cite{lee2005mesh}, multiple segmentation evaluation \cite{golovinskiy2008randomized}, isoline detection \cite{au2011mesh, wang2014spectral}, and edge-based analysis \cite{rodrigues2015contour}. Multiple boundary detection, closed boundary identification and automatic determination of the number of regions are among the most important issues \cite{rodrigues2018part}.

\subsubsection{Skeleton-based segmentation}

Skeleton-based mesh segmentation algorithms complement surface-based methods by capturing the overall structure of shapes through compact 1D representations. Instead of identifying regions with similar geometric properties, these approaches reveal the underlying hierarchical organization and connectivity of shape components. This representation can be constructed using three main approaches: median axis transformation, geometric contraction and Reeb graphs.

The \textbf{Medial Axis Transform}, introduced by \cite{blum1967transformation}, represents the locus of centers of maximal spheres inscribed within a 3D mesh. Some algorithms \cite{brandt1992continuous, amenta2001power} use Voronoi diagrams, while others \cite{wu2006domain} detect non-overlapping maximum spheres. Due to computational complexity, particularly for concave or irregular shapes, applications generally focus on elongated or tubular shapes \cite{mortara2004plumber,mortara2006geometric,sharf2007fly, cornea2007curve}. Recent work has extended its use to machine learning-based approaches \cite{hu2019mat, lin2020seg}.

\textbf{Geometric contraction} reduces shapes to skeletal representations via two different approaches. Volumetric techniques \cite{pudney1998distance,bitter2001penalized,couprie2007discrete} are based on voxel discretization for skeletal extraction. Mesh contraction methods \cite{li2001decomposing, au2008skeleton, damseh2020modeling, ma2024easyskinning} perform iterative smoothing directly on surfaces, preserving topology. The latter method provides greater precision, but may produce the skeleton in several pieces that need to be reconnected.

\textbf{Reeb graphs}, based on Morse theory \cite{milnor1963morse}, characterize shapes by analyzing scalar functions defined on their surfaces, revealing significant structural features through the adjacency of these level-sets. Early techniques used height functions \cite{shinagawa1991surface, xiao2003discrete, KARMAKAR201625}, but their lack of rotational invariance led to inconsistent results. Recent approaches prefer geodesic distance{tierny2007topology,berretti20093d, 9525504} for its invariance properties, despite its higher computational cost. Curvature-based functions \cite{mortara2002affine} are less widespread, but offer invariance to affine transformations and topological changes, with noise managed through adaptive computational radii or filtering.
    
\subsection{Overview and paper organization}

In summary, surface-based and skeletal-based segmentation methods present distinct trade-offs. The former identify regions with similar local features well, but struggle to capture global features, while the latter capture the overall organization of the shape but may lack local detail. This paper proposes to unify these two perspectives using a new method based on the Reeb graph. Unlike previous methods that rely on height or geodesic functions \cite{xiao2003discrete, tierny2007topology, berretti20093d}, this approach supports arbitrary scalar functions while addressing two important issues: the reduction of noisy graphs and the treatment of non-continuous regions. In addition, the Reeb graph serves as an intermediate representation, offering computational efficiency and enabling the rapid generation of multiple segmentations. This flexibility allows the method to be adapted to a variety of applications, from local to global geometric analysis.

The paper proceeds as follows: Section 2 first establishes the theoretical foundations of Morse theory and Reeb graphs. Section 3 then details the method in three phases: Reeb graph construction, topological simplification and continuous region growing. Finally, Section 4 presents two very different applications as case studies. The first is part-based segmentation using the shape diameter function, and the second is chart-based segmentation using the shape index. The paper concludes with future directions.

\section{Theoretical background}

Morse theory and Reeb graphs are two mathematical theories that analyze the topology of shapes through scalar functions defined on them. Morse theory first identifies local topological features by analyzing the scalar values associated with mesh vertices to define critical points. Then, Reeb graphs link these points together by studying the adjacency of the level-sets passing through them.

\subsection{Morse Theory}
\label{Morse Theory} 

Morse theory relates a manifold's differential geometry to its topology by analyzing scalar functions defined on it. Given a manifold $M$, a real-valued function $f$ and $p$ a point on the manifold, if the gradient of $f(p)$ nullifies, then $p$ is called \textit{critical point}. These points are classified as \textit{minimum}, \textit{maximum} and \textit{saddle} respectively. The other points of $M$, where the gradient does not nullify, are called \textit{regular points}.

Let $p$ be a point of $M$ with coordinates $(u,v)$. The Hessian matrix $H_f(p)$, which contains second-order partial derivatives of $f$, is defined as:
\begin{equation}
H_f(p) = 
\begin{bmatrix}
    \frac{\partial^2 f}{\partial u^2}(p) & \frac{\partial^2 f}{\partial u \partial v}(p) \\
    \frac{\partial^2 f}{\partial v \partial u}(p) & \frac{\partial^2 f}{\partial v^2}(p)
\end{bmatrix}
\end{equation}

If $\mathrm{det}(H_f(p)) \neq 0$ then $p$ is a \textit{non-degenerate critical point} and the \textit{Morse index} $k$ of $p$ is obtained by counting the number of negative eigenvalues of the Hessian matrix. For a manifold of dimension 2, the index $k$ of such a point can be used to determine its nature:
\begin{itemize}
    \item if $k = 0$, $p$ is a minimum point;
    \item if $k = 1$, $p$ is a saddle point;
    \item if $k = 2$, $p$ is a maximum point.
\end{itemize}

Continuous Morse theory describes the relationship between differential geometry and topology, but its application to discrete meshes is difficult to implement. Calculating derivatives on piecewise linear surfaces leads to numerical instabilities in the evaluation of the gradients and Hessian matrices needed to classify critical points. Discrete Morse theory, introduced by Banchoff \cite{banchoff1970critique}, addresses these limitations by reformulating fundamental concepts in combinatorial terms. Rather than relying on differential operators, it characterizes critical points by analyzing the neighborhood of vertices, making it naturally suited to triangulated surfaces.

For a vertex $v$ on a 2-manifold surface $S$ and a scalar function $f$ defined on $S$, vertex neighborhood analysis considers two local subgraphs:
\begin{itemize}
\item the \textit{superior link} $Li^+(S,v,f)$ of vertex $v$ is defined as the set of all vertices $v_i$ in the neighborhood $N(v)$ where $f(v_i) > f(v)$, as well as all edges $e_{ij}$ connecting pairs of these vertices. The notation $|Li^+(S,v,f)|$ indicates the number of connected components in the resulting subgraph.
\item the \textit{inferior link} $Li^-(S,v,f)$ follows an analogous definition for vertices where $f(v_i) < f(v)$
\end{itemize}

Particular attention is required for cases where adjacent vertices share identical function values. Ni et al. \cite{ni2004fair} propose to use the Conley index, introduced by Mischaikow and Mrozek \cite{mischaikow2002conley}, which generalizes the Morse index to non-isolated critical points. Unlike the Morse index, which characterizes critical points by the eigenvalues of the Hessian matrix, the Conley index describes the topology of critical regions by the dynamics of their boundaries, making it natural for discrete environments. Thus, their method treats any set of adjacent vertices with equal values as a single vertex and applies the classical link-based characterization of critical points.

\begin{figure}[htb]
    \centering
	\begin{subfigure}[htbp]{0.45\columnwidth}
        \centering
        \begin{tikzpicture}[scale=0.45]
            \centering
    		\node[centernode] (center) at (0, 0) {};
    		\node[nodeMoins] (upleft) at (-2, 1.2) {\faMinus};
    		\node[nodeMoins] (upmiddle) at (0, 2) {\faMinus};
    		\node[nodeMoins] (upright) at (2, 1.2) {\faMinus};
    		\node[nodeMoins] (downleft) at (-2, -1.2) {\faMinus};
    		\node[nodeMoins] (downmiddle) at (0, -2) {\faMinus};
    		\node[nodeMoins] (downright) at (2, -1.2) {\faMinus};
    
    		\path[every node/.style={font=\sffamily\tiny, -}, every edge/.style={draw=black}]
    		(center) edge[-] (upleft)
    		(center) edge[-] (upmiddle)
    		(center) edge[-] (upright)
    		(center) edge[-] (downleft)
    		(center) edge[-] (downmiddle)
    		(center) edge[-] (downright)
    		
    		(upleft) edge[-] (upmiddle)
    		(upmiddle) edge[-] (upright)
    		(upright) edge[-] (downright)
    		(downright) edge[-] (downmiddle)
    		(downmiddle) edge[-] (downleft)
    		(downleft) edge[-] (upleft)
    		;
    	    \end{tikzpicture}
	    \caption{Morse maximum}
	    \label{CriticalMaximum}
    \end{subfigure}
    \hfill
	\begin{subfigure}[htbp]{0.45\columnwidth}
        \centering
    	\begin{tikzpicture}[scale=0.45]
            \centering
            \node[centernode] (center) at (0,0) {};
            
            \node[nodeMoins] (n1) at (-1.5,1.5) {\faMinus};
            \node[nodeMoins] (n2) at (1.5,1.5) {\faMinus};
            \node[nodeMoins] (n3) at (-1.5,-1.5) {\faMinus};
            \node[nodeMoins] (n4) at (1.5,-1.5) {\faMinus};
            
            \path[every node/.style={font=\sffamily\tiny, -}, every edge/.style={draw=black}]
            (n1) edge[-] (center)
            (n2) edge[-] (center)
            (n3) edge[-] (center)
            (n4) edge[-] (center)
            (n1) edge[-] (n3)
            (n2) edge[-] (n4);
	    \end{tikzpicture}
	    \caption{Degenerate maximum}
	    \label{CriticalDegenerateMaximum}
    \end{subfigure}
	\\
	\vspace{2ex}
	\begin{subfigure}[b]{0.45\columnwidth}
        \centering
    	\begin{tikzpicture}[scale=0.45]
		\node[centernode] (center) at (0, 0) {};
		\node[nodePlus] (upleft) at (-2, 1.2) {\faPlus};
		\node[nodePlus] (upmiddle) at (0, 2) {\faPlus};
		\node[nodePlus] (upright) at (2, 1.2) {\faPlus};
		\node[nodePlus] (downleft) at (-2, -1.2) {\faPlus};
		\node[nodePlus] (downmiddle) at (0, -2) {\faPlus};
		\node[nodePlus] (downright) at (2, -1.2) {\faPlus};

		\path[every node/.style={font=\sffamily\tiny, -}, every edge/.style={draw=black}]
		(center) edge[-] (upleft)
		(center) edge[-] (upmiddle)
		(center) edge[-] (upright)
		(center) edge[-] (downleft)
		(center) edge[-] (downmiddle)
		(center) edge[-] (downright)
		
		(upleft) edge[-] (upmiddle)
		(upmiddle) edge[-] (upright)
		(upright) edge[-] (downright)
		(downright) edge[-] (downmiddle)
		(downmiddle) edge[-] (downleft)
		(downleft) edge[-] (upleft)
		;
	\end{tikzpicture}
	\caption{Morse minimum}
	\label{CriticalMinimum}
    \end{subfigure}
    \hfill
	\begin{subfigure}[b]{0.45\columnwidth}
        \centering
    	\begin{tikzpicture}[scale=0.45]
            \node[centernode] (center) at (0,0) {};
            
            \node[nodePlus] (n1) at (-1.5,1.5) {\faPlus};
            \node[nodePlus] (n2) at (1.5,1.5) {\faPlus};
            \node[nodePlus] (n3) at (-1.5,-1.5) {\faPlus};
            \node[nodePlus] (n4) at (1.5,-1.5) {\faPlus};
            
            \path[every node/.style={font=\sffamily\tiny, -}, every edge/.style={draw=black}]
            (n1) edge[-] (center)
            (n2) edge[-] (center)
            (n3) edge[-] (center)
            (n4) edge[-] (center)
            (n1) edge[-] (n3)
            (n2) edge[-] (n4);
	    \end{tikzpicture}
	    \caption{Degenerate minimum}
	    \label{CriticalDegenerateMinimum}
    \end{subfigure}
	\\
	\vspace{2ex}
	\begin{subfigure}[b]{0.45\columnwidth}
        \centering
    	\begin{tikzpicture}[scale=0.45]
		\node[centernode] (center) at (0, 0) {};
		\node[nodePlus] (upleft) at (-2, 1.2) {\faPlus};
		\node[nodePlus] (upmiddle) at (0, 2) {\faPlus};
		\node[nodePlus] (upright) at (2, 1.2) {\faPlus};
		\node[nodeMoins] (downleft) at (-2, -1.2) {\faMinus};
		\node[nodeMoins] (downmiddle) at (0, -2) {\faMinus};
		\node[nodeMoins] (downright) at (2, -1.2) {\faMinus};

		\path[every node/.style={font=\sffamily\tiny, -}, every edge/.style={draw=black}]
		(center) edge[-] (upleft)
		(center) edge[-] (upmiddle)
		(center) edge[-] (upright)
		(center) edge[-] (downleft)
		(center) edge[-] (downmiddle)
		(center) edge[-] (downright)
		
		(upleft) edge[-] (upmiddle)
		(upmiddle) edge[-] (upright)
		(upright) edge[-] (downright)
		(downright) edge[-] (downmiddle)
		(downmiddle) edge[-] (downleft)
		(downleft) edge[-] (upleft)
		;
	\end{tikzpicture}
	\caption{Regular}
	\label{CriticalRegular}
    \end{subfigure}
    \hfill
	\begin{subfigure}[b]{0.45\columnwidth}
        \centering
    	\begin{tikzpicture}[scale=0.45]
		\node[centernode] (center) at (0, 0) {};
		\node[nodePlus] (upleft) at (-2, 1.2) {\faPlus};
		\node[nodeMoins] (upmiddle) at (0, 2) {\faMinus};
		\node[nodePlus] (upright) at (2, 1.2) {\faPlus};
		\node[nodePlus] (downleft) at (-2, -1.2) {\faPlus};
		\node[nodeMoins] (downmiddle) at (0, -2) {\faMinus};
		\node[nodePlus] (downright) at (2, -1.2) {\faPlus};

		\path[every node/.style={font=\sffamily\tiny, -}, every edge/.style={draw=black}]
		(center) edge[-] (upleft)
		(center) edge[-] (upmiddle)
		(center) edge[-] (upright)
		(center) edge[-] (downleft)
		(center) edge[-] (downmiddle)
		(center) edge[-] (downright)
		
		(upleft) edge[-] (upmiddle)
		(upmiddle) edge[-] (upright)
		(upright) edge[-] (downright)
		(downright) edge[-] (downmiddle)
		(downmiddle) edge[-] (downleft)
		(downleft) edge[-] (upleft)
		;
	\end{tikzpicture}
    \caption{Morse saddle}
	\label{CriticalMorse}
    \end{subfigure}
    
	\vspace{2ex}
    \begin{subfigure}[b]{0.45\columnwidth}
        \centering
    	\begin{tikzpicture}[scale=0.45]
		\node[centernode] (center) at (0, 0) {};
		\node[nodePlus] (upleft) at (-2, 1.2) {\faPlus};
		\node[nodeMoins] (upmiddle) at (0, 2) {\faMinus};
		\node[nodePlus] (upright) at (2, 1.2) {\faPlus};
		\node[nodeMoins] (downleft) at (-2, -1.2) {\faMinus};
		\node[nodePlus] (downmiddle) at (0, -2) {\faPlus};
		\node[nodeMoins] (downright) at (2, -1.2) {\faMinus};

		\path[every node/.style={font=\sffamily\tiny, -}, every edge/.style={draw=black}]
		(center) edge[-] (upleft)
		(center) edge[-] (upmiddle)
		(center) edge[-] (upright)
		(center) edge[-] (downleft)
		(center) edge[-] (downmiddle)
		(center) edge[-] (downright)
		
		(upleft) edge[-] (upmiddle)
		(upmiddle) edge[-] (upright)
		(upright) edge[-] (downright)
		(downright) edge[-] (downmiddle)
		(downmiddle) edge[-] (downleft)
		(downleft) edge[-] (upleft)
		;
	\end{tikzpicture}
	\caption{Degenerate saddle}
	\label{CriticalDegenerateSaddle}
    \end{subfigure}
    \hfill
    \begin{subfigure}[b]{0.45\columnwidth}
        \centering
    	\begin{tikzpicture}[scale=0.45]
		\node[centernode] (center) at (0, 0) {};
		\node[nodePlus] (upleft) at (-2, 1.2) {\faPlus};
		\node[nodePlus] (upmiddle) at (0, 2) {\faPlus};
		\node[nodePlus] (upright) at (2, 1.2) {\faPlus};
		\node[nodeMoins] (downleft) at (-2, -1.2) {\faMinus};
		\node[nodeMoins] (downmiddle) at (0, -2) {\faMinus};
		\node[nodeMoins] (downright) at (2, -1.2) {\faMinus};

		\path[every node/.style={font=\sffamily\tiny, -}, every edge/.style={draw=black}]
		(center) edge[-] (upleft)
		(center) edge[-] (upmiddle)
		(center) edge[-] (upright)
		(center) edge[-] (downleft)
		(center) edge[-] (downmiddle)
		(center) edge[-] (downright)
		
		(upleft) edge[-] (upmiddle)
		(upmiddle) edge[-] (upright)
		(upright) edge[-] (downright)
		(downright) edge[-] (downmiddle)
		;
	\end{tikzpicture}
	\caption{Asymmetric saddle}
	\label{CriticalAsymmetricSaddle}
    \end{subfigure}
    \caption{Classification of vertex configurations in discrete Morse theory. }
    \label{CriticalPoints}
\end{figure}
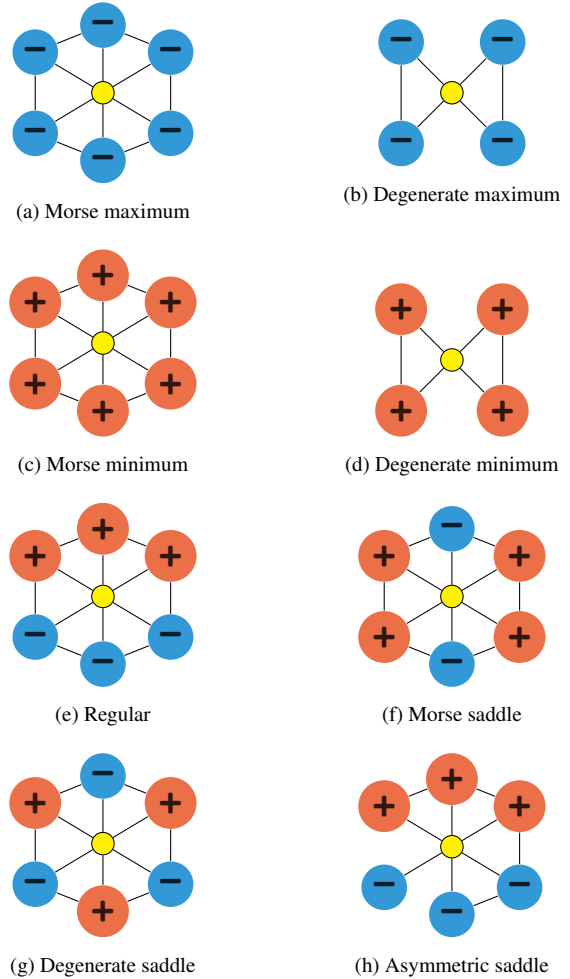

Given a surface $S$, vertices are classified according to their link configurations as follows:
\begin{itemize}
    \item for a vertex $v$ where $|Li^+(S,v,f)| = 0$ and $|Li^-(S,v,f)| = k$ with $k > 0$:
    \begin{itemize}
        \item If $k = 1$: $v$ is a Morse maximum (Figure \ref{CriticalMaximum})
        \item If $k > 1$: $v$ is a degenerate maximum (Figure \ref{CriticalDegenerateMaximum})
    \end{itemize}
    \item for a vertex $v$ where $|Li^+(S,v,f)| = k$ and $|Li^-(S,v,f)| = 0$ with $k > 0$:
    \begin{itemize}
        \item If $k = 1$: $v$ is a Morse minimum (Figure \ref{CriticalMinimum})
        \item If $k > 1$: $v$ is a degenerate minimum (Figure \ref{CriticalDegenerateMinimum})
    \end{itemize}
    \item for a vertex $v$ where $|Li^+(S,v,f)| = |Li^-(S,v,f)| = k$ with $k > 0$:
    \begin{itemize}
        \item If $k = 1$: $v$ is a regular vertex (Figure \ref{CriticalRegular})
        \item If $k = 2$: $v$ is a Morse saddle vertex (Figure \ref{CriticalMorse})
        \item If $k > 2$: $v$ is a degenerate saddle point (Figure \ref{CriticalDegenerateSaddle})
    \end{itemize}
    \item for a vertex $v$ where $|Li^+(S,v,f)| \neq |Li^-(S,v,f)|$ and both $> 0$:
    \begin{itemize}
        \item $v$ is an asymmetric saddle point (Figure \ref{CriticalAsymmetricSaddle})
    \end{itemize}
\end{itemize}

Discrete Morse theory provides a basis for characterizing local topology using critical points, but understanding global structure requires a representation that connects these points.

\subsection{Reeb Graph}
\label{Reeb_theory}

The Reeb graph, introduced by \cite{milnor1963morse}, is a mathematical representation used to analyze the global topological characteristics of a manifold. Based on discrete Morse theory, it connects critical and regular points based on the adjacency of the level-sets associated with the points.

To define these level-sets on a triangulated surface, the scalar function $f : V \rightarrow \mathbb{R}$ defined on the vertices $V$ is extended into the piecewise linear function $f^* : S \rightarrow \mathbb{R}$ by applying the linear interpolation across triangles:
\begin{equation}
f^*(p) := \sum_{v_i \in \sigma}\lambda_i f(v_i), \quad \forall p \in \sigma \in S
\end{equation}
where $\lambda_i$ are the barycentric coordinates of point $p$ in triangle $\sigma$, satisfying:
$\sum_{i}\lambda_i = 1$ and $\lambda_i \geq 0$ for all $i$.
\\

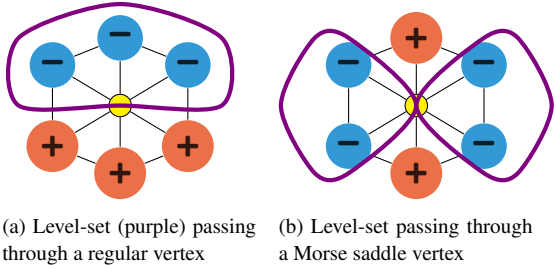
\begin{figure}[htb]
    \centering
    \begin{subfigure}{0.4\columnwidth}
    \begin{tikzpicture}[scale=0.45]
  	    \node[centernode] (center) at (0, 0) {};
	    \node[nodeMoins] (upleft) at (-2, 1.2) {\faMinus};
	    \node[nodeMoins] (upmiddle) at (0, 2) {\faMinus};
	    \node[nodeMoins] (upright) at (2, 1.2) {\faMinus};
	    \node[nodePlus] (downleft) at (-2, -1.2) {\faPlus};
	    \node[nodePlus] (downmiddle) at (0, -2) {\faPlus};
	    \node[nodePlus] (downright) at (2, -1.2) {\faPlus};
	
	    \path[every node/.style={font=\sffamily\tiny, -}, every edge/.style={draw=black}]
	    (center) edge[-] (upleft)
	    (center) edge[-] (upmiddle)
	    (center) edge[-] (upright)
	    (center) edge[-] (downleft)
	    (center) edge[-] (downmiddle)
	    (center) edge[-] (downright)
	
	    (upleft) edge[-] (upmiddle)
	    (upmiddle) edge[-] (upright)
	    (upright) edge[-] (downright)
	    (downright) edge[-] (downmiddle)
	    (downmiddle) edge[-] (downleft)
	    (downleft) edge[-] (upleft)
	    ;
        \draw [violet, ultra thick] plot [smooth cycle] coordinates {(0,0) (3,0) (3, 2) (0,3) (-3, 2)  (-3,0)};
    \end{tikzpicture}
    \caption{Level-set (purple) passing through a regular vertex}
    \label{fig:levelSet_regular}
    \end{subfigure}
    \hspace{1ex}
    \begin{subfigure}{0.4\columnwidth}
    \begin{tikzpicture}[scale=0.45]
	    \node[centernode] (center) at (0, 0) {};
    	\node[nodeMoins] (upleft) at (-2, 1.2) {\faMinus};
	    \node[nodePlus] (upmiddle) at (0, 2) {\faPlus};
	    \node[nodeMoins] (upright) at (2, 1.2) {\faMinus};
	    \node[nodeMoins] (downleft) at (-2, -1.2) {\faMinus};
	    \node[nodePlus] (downmiddle) at (0, -2) {\faPlus};
	    \node[nodeMoins] (downright) at (2, -1.2) {\faMinus};
	
    	\path[every node/.style={font=\sffamily\tiny, -}, every edge/.style={draw=black}]
	    (center) edge[-] (upleft)
	    (center) edge[-] (upmiddle)
	    (center) edge[-] (upright)
	    (center) edge[-] (downleft)
	    (center) edge[-] (downmiddle)
	    (center) edge[-] (downright)
	
    	(upleft) edge[-] (upmiddle)
	    (upmiddle) edge[-] (upright)
	    (upright) edge[-] (downright)
	    (downright) edge[-] (downmiddle)
	    (downmiddle) edge[-] (downleft)
	    (downleft) edge[-] (upleft)
	    ;
	    \draw [violet, ultra thick] plot [smooth cycle] coordinates {(0,0) (2,-2) (3, -2) (4,0) (3, 2)  (2, 2)};
	    \draw [violet, ultra thick] plot [smooth cycle] coordinates {(0,0) (-2,-2) (-3, -2) (-4,0) (-3, 2)  (-2, 2)};
    \end{tikzpicture}
    \caption{Level-set passing through a Morse saddle vertex}
    \label{fig:levelSet_saddle}
    \end{subfigure}
    \caption{Level-set configurations: regular vertex (a) and Morse saddle vertex (b) with their isolines.}
    \label{fig:levelSets}
\end{figure}

A level-set of value $l$ for $f^*$ on $S$ consists of all points $x$ such that $f^*(x) = l$. The connected components of a level-set are called \textit{isolines}. For a vertex $v$ of $S$, the shape of the isoline passing through this vertex is determined by its type:
\begin{itemize}
\item if $v$ is a minimum or maximum vertex, then the isoline degenerates to a point;
\item if $v$ is a regular vertex, then the isoline forms a closed line, as illustrated in Figure \ref{fig:levelSet_regular};
\item if $v$ is a saddle vertex, then the isoline consists of multiple closed lines intersecting at $v$, as shown in Figure \ref{fig:levelSet_saddle}.
\end{itemize}

Thus, nodes of the Reeb graph are formed by the level-sets that pass through each point of the surface, and edges between nodes are determined by the adjacency of their corresponding level-sets. Formally, the Reeb graph of $f^*$ is defined as the quotient space under the equivalence relation:
\begin{equation}
(x,f^*(x)) \sim (y,f^*(y)) \Leftrightarrow f^*(x) = f^*(y) = l
\end{equation}

The topology of a node $u$ of the graph is described by analyzing its degree $d(u)$, its in-degree $d^-(u)$ and its out- degree $d^+(u)$ as follows:
\begin{itemize}
\item if $u$ is a regular node, then $d^-(u) = d^+(u) = 1$;
\item if $u$ is a minimum node, then $d^-(u) = 0$ and $d^+(u) \geq 1$, with degenerate minima having $d(u) > 1$;
\item if $u$ is a maximum node, then $d^-(u) \geq 1$ and $d^+(u) = 0$, with degenerate maxima having $d(u) > 1$;
\item if $u$ is a saddle vertex, then $d^-(u) \geq 1$,  $d^+(u) \geq 1$ and $d(u) \geq 3$, with degenerate saddles having $d(u) > 3$.
\end{itemize}

\begin{figure}[htb]
     \centering
     \begin{subfigure}{0.48\columnwidth}
     \centering
         \includegraphics[width=1\columnwidth,keepaspectratio]{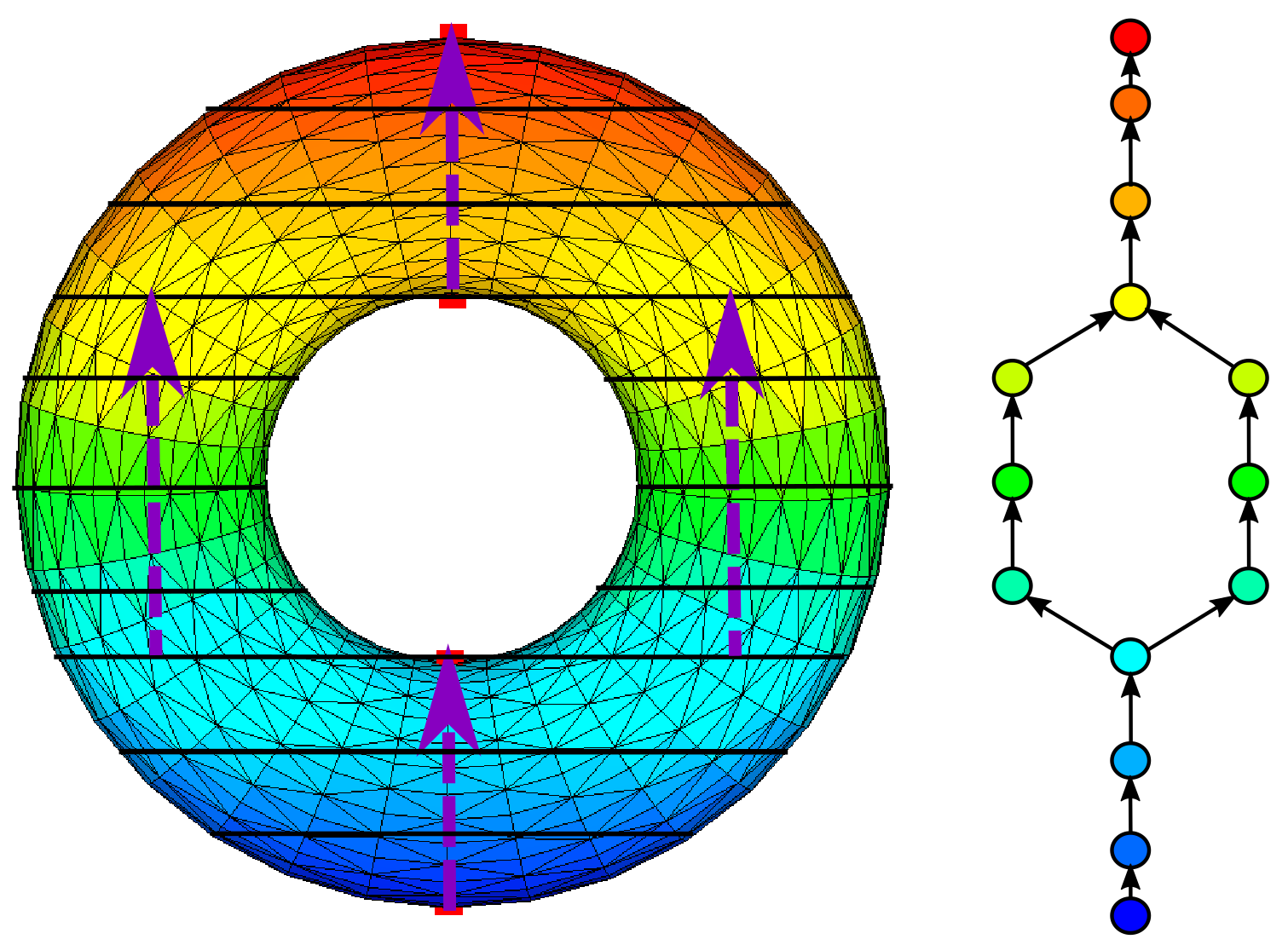}
         \caption{Y-direction height function with two Morse saddle points creating a cyclic Reeb graph}
         \label{fig:reebTorus}
     \end{subfigure}
     \hfill
     \begin{subfigure}{0.48\columnwidth}
     \centering
         \includegraphics[width=1\columnwidth,keepaspectratio]{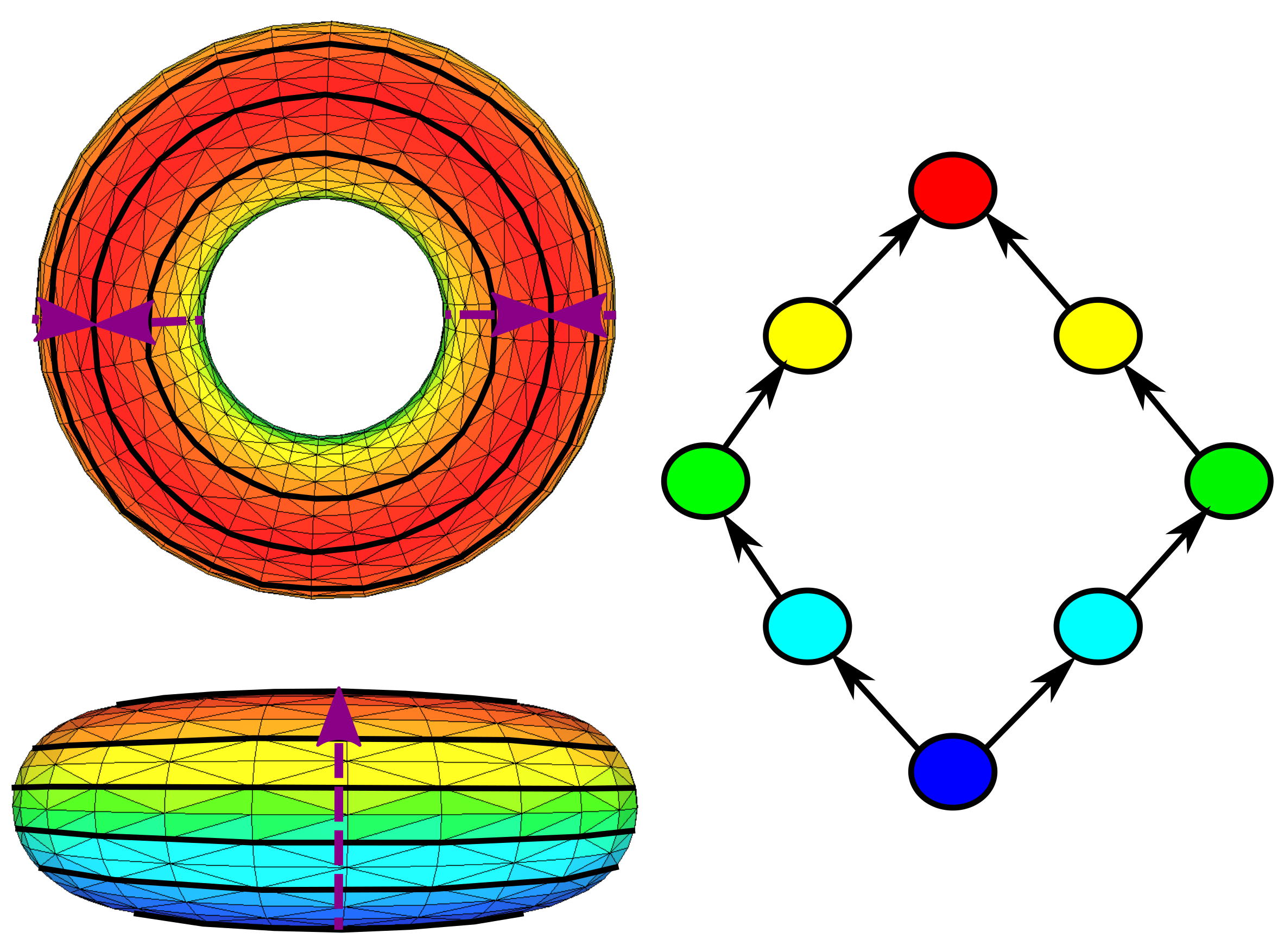}
         \caption{Z-direction height function with two degenerate extrema creating a cyclic Reeb graph}
         \label{fig:reebTorusZ}
     \end{subfigure}
        \caption{Level-sets and Reeb graphs of a torus, comparing Morse (a) and non-Morse (b) height functions.}
        \label{fig:reebGraphTorus}
\end{figure}

The construction and structure of Reeb graphs is illustrated by Figure \ref{fig:reebGraphTorus} showing two scalar functions applied to the same torus mesh. The $Y$-direction height function, shown in Figure \ref{fig:reebTorus}, produces four non-degenerate critical points: a minimum (blue), a maximum (red), and two saddles (cyan, yellow). This is a Morse function and its cyclic Reeb graph with two branches reflects the torus's fundamental topology. The $Z$-direction height, illustrated in Figure \ref{fig:reebTorusZ}, generates two degenerate critical points: a minimum (blue) and a maximum (red). This is a non-Morse function and its Reeb graph is purely cyclic. These cases highlight how Reeb graphs provide a global topological representation of shapes through their critical points, which make them suitable for mesh segmentation when the scalar function represents its geometrical features.

\section{Methodology}

The segmentation algorithm consists of three phases. The first phase constructs the Reeb graph from a user-defined scalar function using an enhanced topological skeleton approach. The second phase simplifies the graph topology by eliminating irrelevant Morse and degenerate critical points. The final phase generates continuous segments through an adaptive region-growth process, balancing local mesh connectivity with global topological constraints of the Reeb Graph according to application requirements.

\subsection{Reeb graph Construction}
\label{sect:reeb_graph_initialization}

Reeb graphs of meshes can be constructed using three types of algorithms, which differ in their handling of computational costs, topological accuracy and sensitivity to noise.

The \textbf{level diagram} \cite{lazarus1999level, hetroy2003topological} algorithms use a sampling step on the scalar function to generate level sets on the mesh. The Reeb graph is then constructed by analyzing the topological changes of successive level sets. Originally designed for zero-genus surfaces, this method has also been extended to handle more complex topologies. However, the use of a sampling parameter can impact either its accuracy or its computational cost.

The \textbf{extended Reeb graphs} algorithms \cite{biasotti2000shape,attene2003shape} sample the scalar function into regions to represent level-sets. The graph is then constructed by analyzing the neighborhood between regions, and an explicit topological evaluation decomposes the region into subregions if it contains a cycle. However, its dependence on interval size can result in the loss of important features. The \textbf{multi-resolution Reeb graph} algorithm \cite{hilaga2001topology} algorithms attempt to overcome these limitations by proposing multi-scale sampling of the scalar function. 

\begin{figure}[htb]
    \centering
	\includegraphics[width=1\columnwidth]{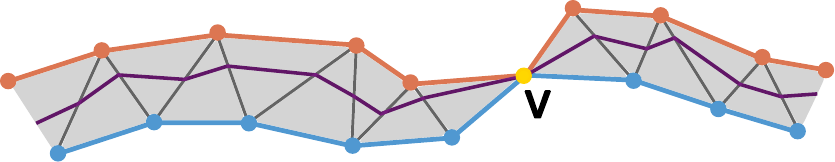}
	\caption{Strip of triangles surrounding the level-set (purple) of a regular point (yellow). Discrete contours at the top (red) and bottom (blue) represent this level-set.}
	\label{DiscretIsoline}
\end{figure}

The \textbf{enhanced topological skeleton} algorithms \cite{tierny2009partial,brandolini2012computing, parsa2012deterministic} take a different approach by defining the concept of discrete contours to approximate level sets instead of sampling. Illustrated in Figure \ref{DiscretIsoline}, a level set (purple) passing through a regular vertex is bounded by discrete contours above (red) and below (blue). Thus, each vertex in the mesh is characterized by these discrete contours and the Reeb graph is constructed as their adjacency graph. This algorithm provides a complete characterization of the local topology without the need for parameter tuning. In addition, the extension proposed by Para \cite{parsa2012deterministic} allows construction with an optimal complexity of $O(n \log n)$ for a mesh of $n$ vertices. However, it is highly sensitive to local variations in the scalar function, including those induced by noise or minor geometric fluctuations.

The enhanced topological skeleton algorithm proposed by Parsa \cite{parsa2012deterministic} is chosen for mesh segmentation. This choice is motivated firstly by the topological reliability of the Reeb graph constructed and by its optimal computational cost. But more importantly, the algorithm provides a direct correspondence between mesh and Reeb graph: each vertex (or contiguous set of vertices with the same scalar value) is associated with a specific graph node. This property establishes a natural connection between the two topological representations. However, its implementation requires a data structure adapted for use within this context. The graph representation is optimized by retaining only critical nodes and their connections and the regular nodes between the critical nodes are stored as sorted lists within the edges, thus preserving the complete topology.  Specifically, for each edge $e$ connecting critical nodes $c_1$ and $c_2$, the algorithm maintains an ordered sequence ${u_1, ..., u_k}$ of regular nodes such that:
$f(c_1) \leq f(u_1) \leq ... \leq f(u_k) \leq f(c_2)$ or $f(c_1) \geq f(u_1) \geq ... \geq f(u_k) \geq f(c_2)$.

\subsection{Topological Simplification}

The Reeb graph constructed with the improved topological skeleton algorithm maintains a direct correspondence between the mesh and the graph structure. However, when working with geometric scalar functions such as curvature measures, this initial graph often contains many critical nodes, both simple and degenerate, which reflect local surface variations rather than significant shape features.

The standard approach to topological simplification uses persistence-based cancellation, which removes pairs of critical nodes according to their relative importance. But this method assumes only non-degenerate (Morse) critical nodes. The degenerate nodes must then be unfolded beforehand into several simple nodes \cite{Edelsbrunner2010}. Nevertheless, this operation breaks the correspondence between mesh vertices and graph nodes important for mesh segmentation, as these new nodes do not correspond to actual mesh vertices.

In this section, we propose a new critical node cancellation algorithm that handles both simple and degenerate critical nodes and preserves the correspondence between the surface and the Reeb graph. This method also combines different metrics to identify which pair of critical nodes should be deleted, rather than relying only on persistence.

\subsubsection{Evaluation Metrics}

The persistence metric $P(e)$ quantifies the topological importance of features by measuring the difference in scalar values between nodes $n_1$ and $n_2$ connected by edge $e$:
\begin{equation}
P(e) = |f(n_1) - f(n_2)|
\end{equation}
High persistence values indicate more significant features in the topology of the scalar function than low persistence values.

The area ratio $R_a(e)$ evaluates the geometrical importance of all vertices located in the edge $e$ connecting $n_1$ and $n_2$ using a scale-invariant surface area measurement:
\begin{equation}
R_a(e) = \frac{\sum_{v \in V(e)} A(v) + \sum_{v \in N(e)} \frac{A(v)}{\mathrm{deg}(v)}}{A_{total}}
\end{equation}
where:
\begin{itemize}
    \item $V(e)$ represents the set of vertices along edge $e$
    \item $N(e)$ comprises vertices in nodes $n_1$ and $n_2$
    \item $A(v)$ denotes the area associated with vertex $v$
    \item $\mathrm{deg}(v)$ indicates the number of edges incident to the node containing $v$
    \item $A_{total}$ represents the total surface area of the mesh
\end{itemize}
Low area ratio values indicate geometrically small regions that may represent noise or minor surface variations in the mesh geometry.

Given two user-defined thresholds $\tau_a$ and $\tau_p$, an edge $e$ is considered a candidate for cancellation if at least one of its nodes is an extremum and $R_a(e) < \tau_a \text{ or } P(e) < \tau_p$. However, if there are several edges between the same pair of nodes, then these edges are not candidates as they represent important topological features such as handles and tunnels in the underlying mesh. Thus, the user has greater control over the features they wish to preserve than with persistence alone. Additional metrics can be added if needed for the user's application.

\subsubsection{Cancellation Algorithm}

\begin{figure*}[b]
    \centering
    \begin{subfigure}{0.48\columnwidth}
        \centering
        \includegraphics[width=1\columnwidth]{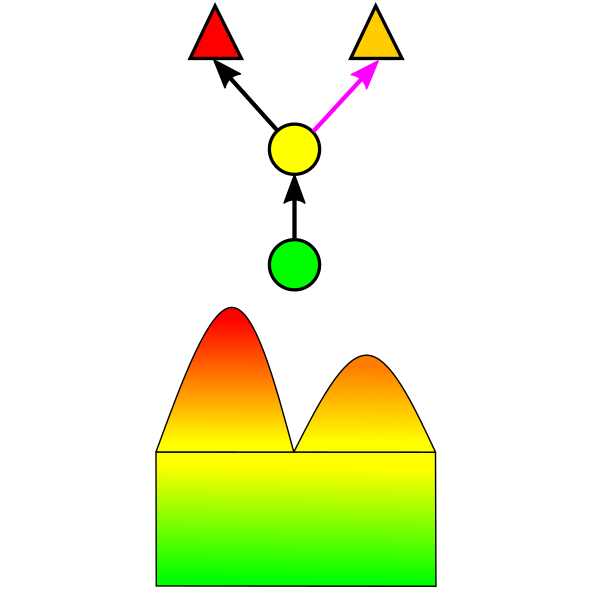}
        \caption{Initial: Morse saddle connected to simple extremum}
        \label{fig:a}
    \end{subfigure}
    \hfill
    \begin{subfigure}{0.48\columnwidth}
        \centering
        \includegraphics[width=1\columnwidth]{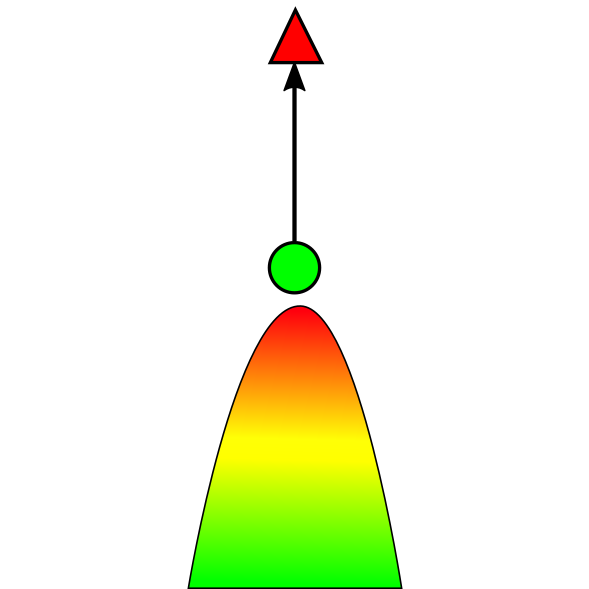}
        \caption{After: saddle becomes regular node within edge}
        \label{fig:b}
    \end{subfigure}
    \hfill
    \begin{subfigure}{0.48\columnwidth}
        \centering
        \includegraphics[width=1\columnwidth]{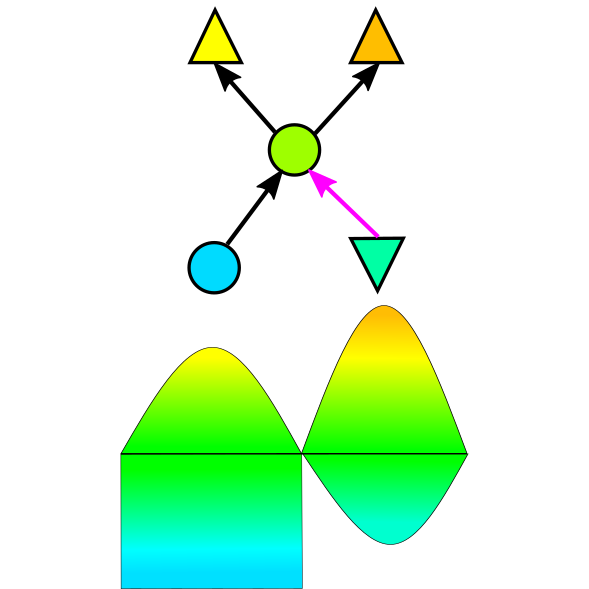}
        \caption{Initial: degenerate saddle connected to simple extremum}
        \label{fig:c}
    \end{subfigure}
    \hfill
    \begin{subfigure}{0.48\columnwidth}
        \centering
        \includegraphics[width=1\columnwidth]{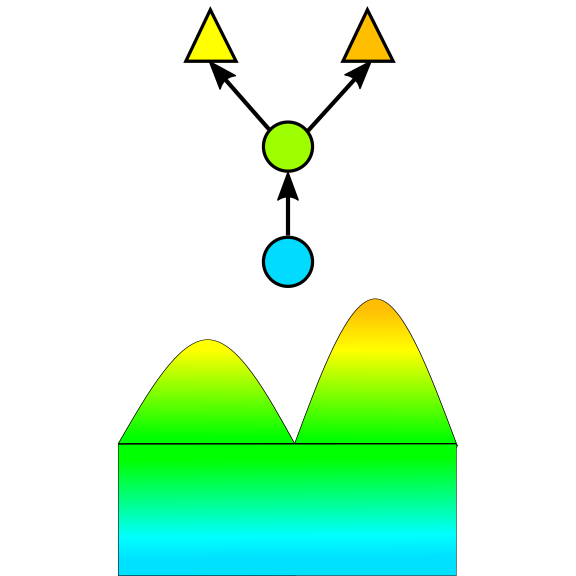}
        \caption{After: saddle preserved with adjusted value}
        \label{fig:d}
    \end{subfigure}
    \begin{subfigure}{0.48\columnwidth}
        \centering
        \includegraphics[width=1\columnwidth]{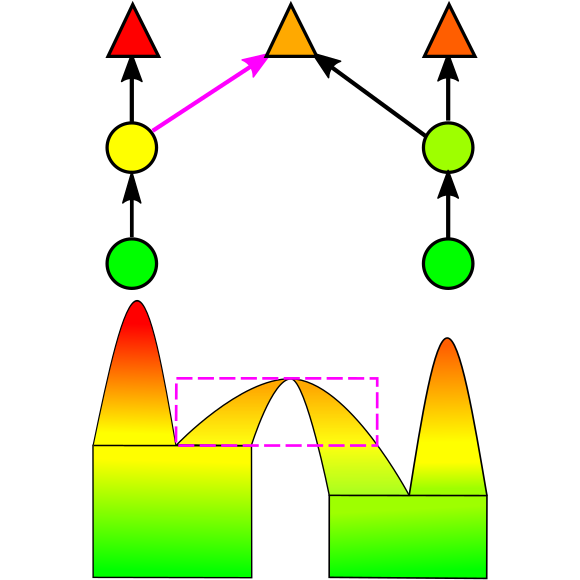}
        \caption{Initial: Morse saddle connected to degerate extremum}
        \label{fig:e}
    \end{subfigure}
    \hfill
    \begin{subfigure}{0.48\columnwidth}
        \centering
        \includegraphics[width=1\columnwidth]{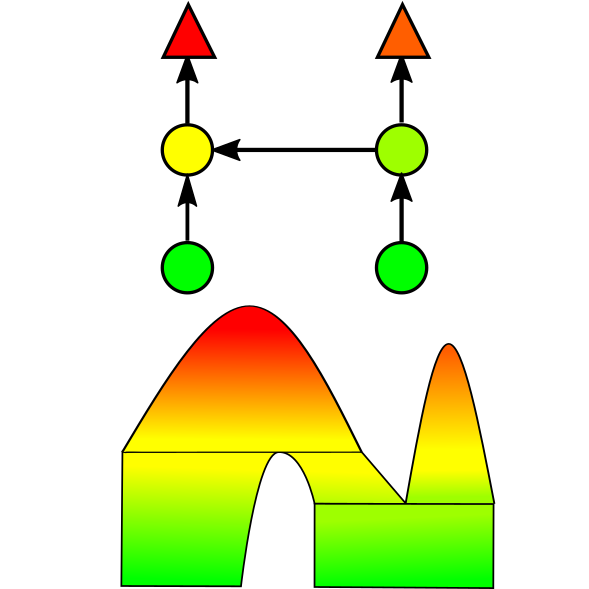}
        \caption{After: saddle preserved with adjusted value}
        \label{fig:f}
    \end{subfigure}
    \hfill
    \begin{subfigure}{0.48\columnwidth}
        \centering
        \includegraphics[width=1\columnwidth]{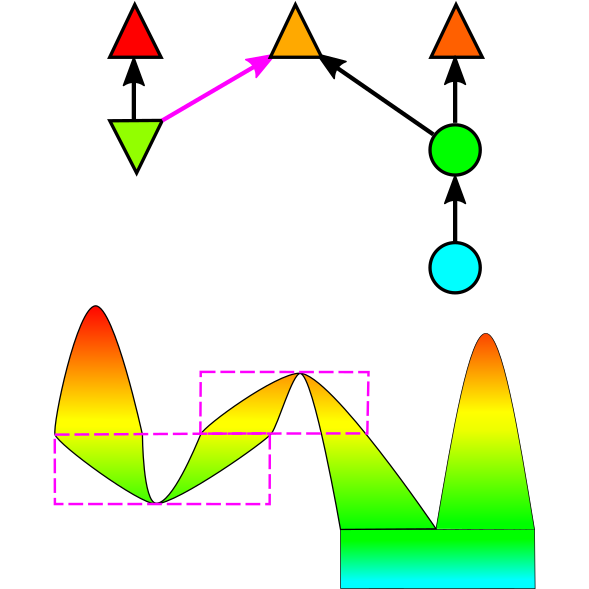}
        \caption{Initial: two connected degenerate extrema}
        \label{fig:g}
    \end{subfigure}
    \hfill
    \begin{subfigure}{0.48\columnwidth}
        \centering
        \includegraphics[width=1\columnwidth]{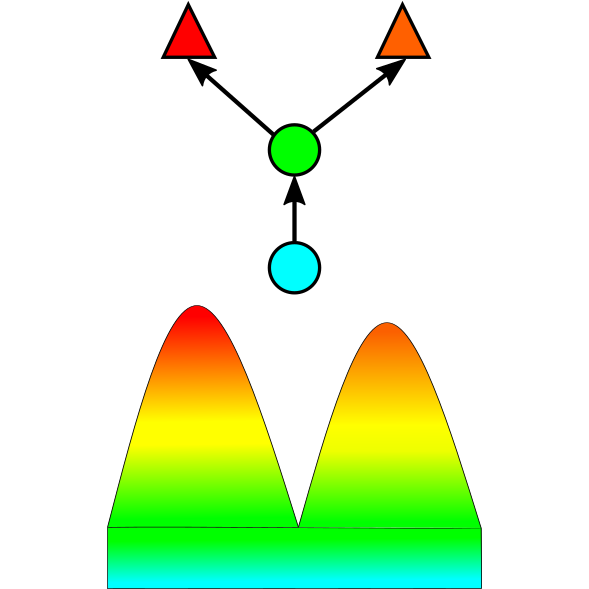}
        \caption{After: extrema merged into regular node within edge}
        \label{fig:h}
    \end{subfigure}
    \caption{Critical node cancellation operations illustrated through Reeb graphs (top) and continuous function representations (bottom). Node colors indicate scalar values (blue: minimum to red: maximum). Critical nodes shown as triangles (extrema) or circles (saddle nodes). Purple highlights mark cancellation regions and affected edges.}
    \label{fig:cancellation_process}
\end{figure*}

Candidate edges are processed iteratively using a self-balancing binary search tree, ordered according to persistence values. This ensures that the least important features are eliminated from the most important, as in the case of persistence diagram filtering. The algorithm then applies one of two cancellation operations based on the Morse classification of the edge nodes:
\begin{description}
    \item[Saddle-Extremum Cancellation] For edges connecting a saddle node and an extremum node, the saddle is preserved while the extremum and the edge are removed. 
    \item[Extremum-Extremum Cancellation] For edges connecting two extrema, both nodes and their edges are removed and replaced by a new node. This node is assigned the median scalar value of the original extrema. 
\end{description}
For both operations, when an edge is removed, the vertices associated with its regular nodes are associated with the preserved or created node, and their scalar value becomes that of this node. The same applies to the removed extremum. In addition, if the extremum is degenerate, the connectivity of the graph must be maintained by reconnecting its other incident edges to the preserved or created node. To preserve the order of the regular nodes in these edges, as described in the section \ref{sect:reeb_graph_initialization}, all those that have a scalar value higher (or lower depending on the direction of the scalar value in the edge) than that of the newly connected node are removed and associated with this node. Their scalar value is modified to reflect the change. Finally, if the preserved or created node becomes a regular node after the incident edges have been reconnected, then its two incident edges are merged.  It is thus placed between the two merged lists of regular nodes in the new edge to maintain the appropriate order based on their scalar values. At the end of the operation, the metric values of the modified edges are re-evaluated as potential candidates. Iterations continue until no valid candidates remain.

To illustrate cancellation operations, four representative scenarios are shown in Figure \ref{fig:cancellation_process}. Each scenario shows the state before (left) and after (right) a cancellation operation applied to a candidate edge of a Reeb graph. To better visualize the changes, two complementary representations are displayed for each state: a Reeb graph (top) and the corresponding continuous function (bottom). Critical nodes are represented by triangles for extrema and circles for saddles and colors indicate scalar values ranging from blue (minimum) through green and yellow to red (maximum). 

\begin{itemize} 
    \item Scenario (a-b) illustrates saddle-extremum cancellation when both nodes are non-degenerate. The operation removes the extremum and its edge, adjusting the scalar values of the vertices to the value of the saddle to preserve continuity. The saddle node, now having exactly one incoming and one outgoing edge, becomes regular and is integrated into the merged edge. 

    \item Scenario (c-d) illustrates saddle-extremum cancellation when the saddle is degenerate. The operation removes the extremum and its edge, adjusting he scalar values of the vertices to the value of the saddle to preserve continuity. The degenerated saddle maintains its classification.
    
    \item Scenario (e-f) illustrates saddle-extremum cancellation when the extremum is degenerate. The operation removes the extremum and its edge and reconnect the incident edge to saddle. The saddle value is propagated throughout the vertices in affected region (purple box) to preserve continuity.
    
    \item Scenario (g-h) illustrates extremum-extremum cancellation when both extrema are degenerate. The operation merges both nodes into a single new node with an intermediary scalar value and reconnect incident edges to this node. This value is propagated throughout the vertices in affected regions (purple boxes) to preserve continuity. Since the resulting node is regular with single incoming and outgoing edges, it is integrated into the merged edge. 
\end{itemize}

Thus, this algorithm simplifies the topology of the Reeb graph while handling both simple and degenerate critical nodes. Vertex-graph correspondence, important for applications such as segmentation, is preserved by adjusting the scalar function values on the mesh vertices according to the simplified topology. As the implementation uses a self-balancing binary search tree to manage candidate edges, the algorithm achieves a complexity of $O(e \log e)$, where $e$ represents the total number of candidate edges. Furthermore, by processing edges according to their persistence values, the algorithm guarantees deterministic execution analogous to persistence diagram filtering.

\subsection{Region Growing}

\begin{figure*}[b]
     \centering
     \begin{subfigure}{0.48\textwidth}
     \centering
          \includegraphics[width=0.65\textwidth]{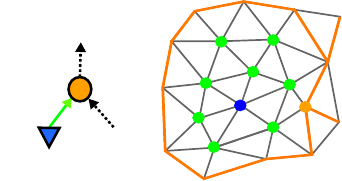}
         \caption{A Reeb graph edge (in green) with a minimum node (in blue) and a saddle node (in orange). The discrete outline of both nodes and the set of regular vertices in the edge are highlighted on the mesh.}
         \label{fig:continuiteA}
     \end{subfigure}
     \hfill
     \begin{subfigure}{0.48\textwidth}
     \centering
         \includegraphics[width=1\textwidth]{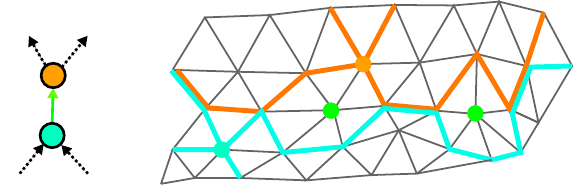}
         \caption{A Reeb graph edge (in green) with two saddle nodes (in cyan and orange). The discrete outline of both nodes and the set of regular vertices in the edge are highlighted on the mesh.}
         \label{fig:continuiteB}
     \end{subfigure}
        \caption{Two edges of a Reeb graph and their discrete local representation on the mesh.}
        \label{fig:continuite}
\end{figure*}

Traditional segmentation methods based on the Reeb graph \cite{tierny2007topology, berretti20093d}, use the critical nodes of the graph to form the regions on the mesh. However, these methods, based solely on the connectivity and values of the graph, do not guarantee their continuity. This problem, less obvious for height or geodesic functions, is nevertheless present for functions with many local variations, such as those based on curvature. It is caused by the lack of correlation between the connectivity of vertices in the reeb graph and that of vertices in the mesh.

Figure \ref{fig:continuite} illustrates this problem of non-continuity in the propagation of regions from critical nodes. A region formed by the set of vertices located between an extremum and a saddle is always continuous (Figure \ref{fig:continuiteA}). This is because, since the extremum is a vertex or a set of contiguous vertices, the vertex of the discrete contour adjacent to it will also be adjacent to one of these vertices on the mesh. However, a region formed by the set of vertices between two saddles is not always continuous (Figure \ref{fig:continuiteB}). Indeed, two vertices adjacent to each other in the Reeb graph by their discrete contour are not necessarily adjacent on the mesh.  

We therefore propose a region growing algorithm that starts by initializing regions from the vertices present in the extrema of the graph. Saddles are not used to create regions because, depending on the scalar function used, they do not describe distinct features of the shape. For example, a function based on curvature may have several saddle nodes with successive close values, which cannot generate a discriminating region of the shape. 

Other graph nodes with their vertices are candidates for growing a region if they are at least adjacent to it on the mesh to ensure its continuity. This adjacency between a node $u$ and a region $R$ is determined by checking whether at least one vertex of $u$ is adjacent to a vertex of $R$ on the mesh:
\begin{equation}
\text{isAdjacent}(u,R) = \exists v \in V(u), v_r \in V(R) \mid v_r \in \text{Adj}(v)
\end{equation}
where:
\begin{itemize}
\item $V(u)$ represents the set of mesh vertices associated with Reeb graph node $u$
\item $V(R)$ represents all mesh vertices currently assigned to region $R$
\item $\text{Adj}(v)$ represents vertices adjacent to $v$ in the mesh
\end{itemize}

This adjacency is not sufficient to propagate regions according to shape features, so we propose two cost functions to be chosen by the user. The first is a standard cost based solely on the geometrical features of the scalar function. The second is a Reeb graph-constrained cost, similar to the first, but incorporating a region adjacency constraint in the Reeb graph to support topological features.

\begin{description}

    \item[Standard cost] A node $u$ becomes a candidate for a region $R$ if $\text{isAdjacent}(u,R)$ is true. Its cost is defined by :
    \begin{equation}
        \text{Cost}_{\text{std}}(u, R) = |\text{mean}(f(R)) - f(u)|
    \end{equation}
    where $\text{mean}(V(R))$ represents the area-weighted average of the scalar values of the vertices in $R$. This cost allows regions to grow freely on the mesh surface according to geometrical features.\\
    
    \item[Reeb graph-constrained cost] A node $u$ becomes a candidate for a region $R$ if $\text{isAdjacent}(u,R)$ is true and if at least one node adjacent to $u$ in the Reeb graph is assigned to any region. Its cost is defined by :
    \begin{equation}
        \text{Cost}_{\text{graph}}(u, R) = |\text{mean}(f(R)) - f(u)| + \lambda \times \mathbb{1}_{{|\text{Unassigned}(u)| > 1}}
    \end{equation}
    where $\lambda$ represents a user-defined parameter and ($| \text{Unassigned} (u)| > 1$) is an indicator function that applies the growth penalty at saddle nodes if more than one branch is not assigned to a region. Once all but one branch of a saddle node has been assigned to regions, its cost to each region is no longer penalized, and it will be assigned to the region that minimizes it. This cost allows regions to grow from extrema on the reeb graph according to geometrical and topological features, while maintaining region continuity on mesh.

\end{description}

The algorithm grows regions by choosing the node-region pair that minimizes one of the above functions. The implementation uses a self-balancing binary search tree that sorts candidate nodes based on their respective cost. In addition, when a node is assigned to a region, the cost of nearby candidates is re-evaluated to guarantee order consistency, and new nodes that now validate the conditions are added. This implementation allows the algorithm to achieve $O(n \log n)$ time complexity for a mesh with $n$ vertices.

\section{Results and Discussion}

    \begin{figure*}[hb]
        \centering
        \begin{subfigure}{0.31\textwidth}
            \centering
            \includegraphics[width=0.8\textwidth]{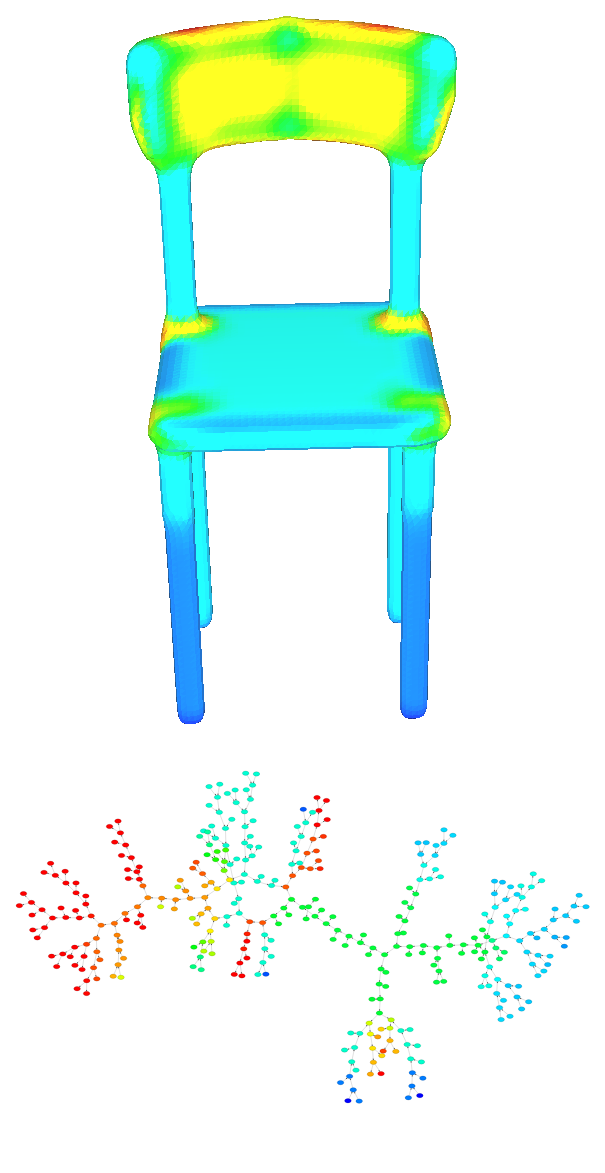}
            \caption{SDF mapped on chair (top) with initial Reeb graph (bottom)}
            \label{Chair_reeb_graph}
        \end{subfigure}
        \hfill
        \begin{subfigure}{0.31\textwidth}
            \centering
            \includegraphics[width=0.8\textwidth]{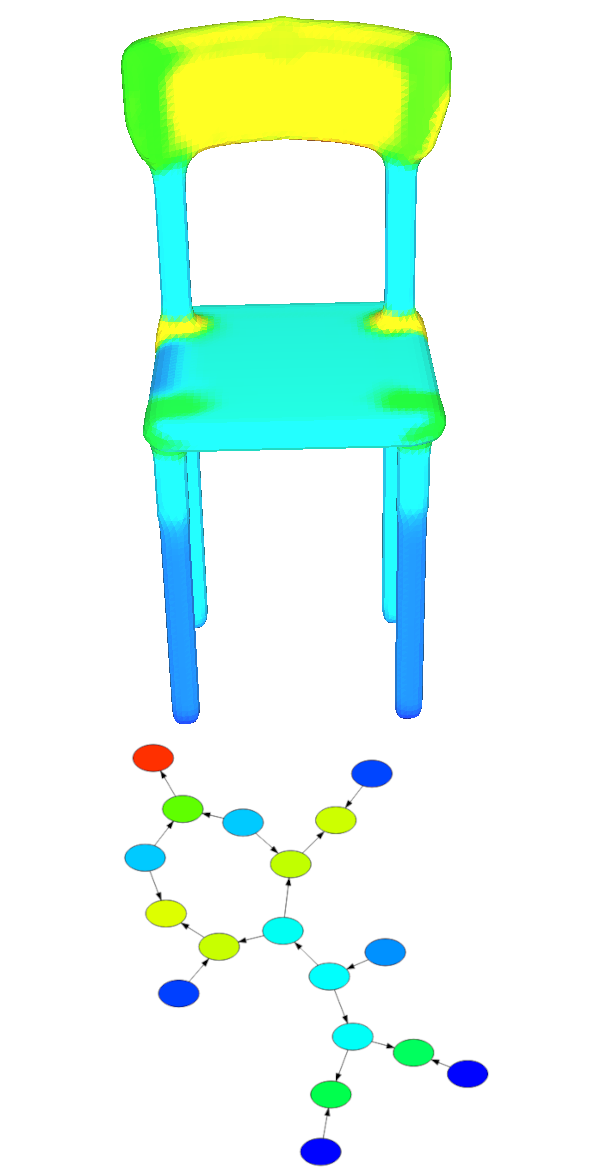}
            \caption{Filtered SDF on chair (top) and simplified Reeb graph (bottom)}
            \label{Chair_cancellation}
        \end{subfigure}
        \hfill
        \begin{subfigure}{0.31\textwidth}
            \centering
            \includegraphics[width=0.8\textwidth]{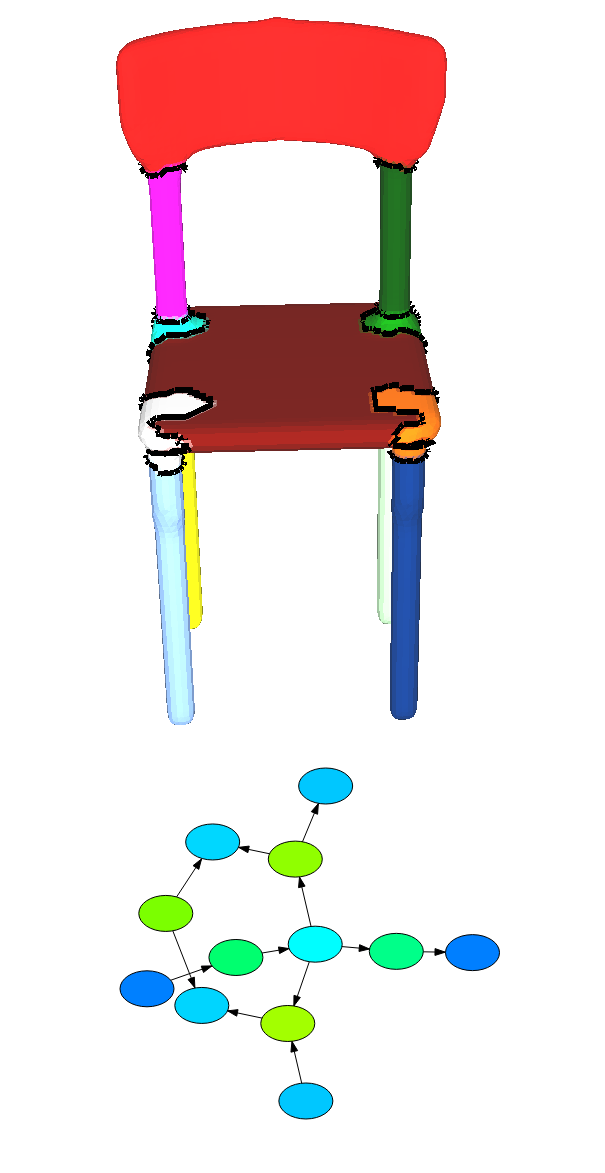}
            \caption{Regions mapped on chair (top) and corresponding region adjacency graph (bottom).}
            \label{chair_segmentation}
        \end{subfigure}
        \caption{Three steps of the segmentation process on a chair model using Shape Diameter Function: initial construction, topological simplification, and region growing.}
        \label{Chair_example}
    \end{figure*}
    
    To validate this segmentation method, we present case studies of two different applications. The first application is a part-based decomposition using the Shape Diameter Function (SDF) on models from the Princeton Benchmark dataset \cite{chen2009benchmark}. In this study, we illustrate the different phases of the method, compare the impact of growth costs on results and analyze both segmentation quality and computational performance. The second application is a chart-based decomposition using Shape Index on a tile extracted from the asteroid Vesta \cite{Vesta}. In this study, we illustrate how area ratio parameters control feature detection and highlight the importance of topological features when working with curvature-based functions.
    
	\subsection{Shape Diameter Function}

    This first application is a part-based decomposition using the Shape Diameter Function (SDF), a fundamental volumetric shape descriptor introduced by \cite{gal2007} and further developed by \cite{shapira2008consistent}. The SDF is a scalar function that enables the identification of semantically significant parts through local thickness measurements \cite{shapira2010}. These values, close to the median axis transformation, provide a local characterization of the shape, invariant to pose transformations.
    
    The SDF values are calculated using a ray-casting methodology that evaluates the thickness of the local volume. For each vertex $v_i$ on the mesh surface, the algorithm constructs a cone centered on the inward-normal vector $n_i$:
    \begin{equation}
        SDF(v_i) = \frac{\sum_{j=1}^{N} w_j d_j}{\sum_{j=1}^{N} w_j}
    \end{equation}
    where $d_j$ represents the ray-intersection distances and $w_j$ denotes the angular weights. For implementation purposes, this study utilizes the SDF computation provided by the CGAL mesh library \cite{cgalSDF} with default parameters.
    
    \subsubsection{Example on a Chair Model}

    Figure \ref{Chair_example} illustrates the three phases of the method through a decomposition based on the parts of a chair model. Each step is visualized with the 3D model (top) and the corresponding graphical representation (bottom). Figure \ref{Chair_reeb_graph} illustrates the initial phase of Reeb graph construction. The distribution of values for the shape diameter function is represented on the model by colors ranging from low (blue) to high (red). The Reeb graph constructed from this function contains numerous critical nodes, reflecting the local variations inherent in volumetric measurements of the shape.

    The topological simplification phase, illustrated in Figure \ref{Chair_cancellation}, applies filter parameters of 6.0 for area ratio and 10 for persistence. The simplified Reeb graph shows several degenerate extrema characterizing the intersections between the seat and the chair back, as well as the junctions of the feet with the seat. This illustrates the importance of managing degenerate nodes in cancellation. The scalar function applied to the mesh was smoothed in conjunction with the graph operations, removing local variations while preserving the main features. 
    
    Figure \ref{chair_segmentation} shows the result of the region growth phase in identifying semantically meaningful components. The distinct colored regions clearly delineate the chair's main structural elements (back, seat, legs), while the adjacency graph of the regions indicates their spatial relationships. In particular, the graph preserves the cycle that represents the chair's characteristic topology, thus showing the method's ability to preserve global topological features.
    
    \subsubsection{Comparison of Region Growing Costs}
    
\begin{figure}[htb]
        \centering
        \begin{subfigure}{0.5\textwidth}
            \centering
            \includegraphics[width=1\textwidth]{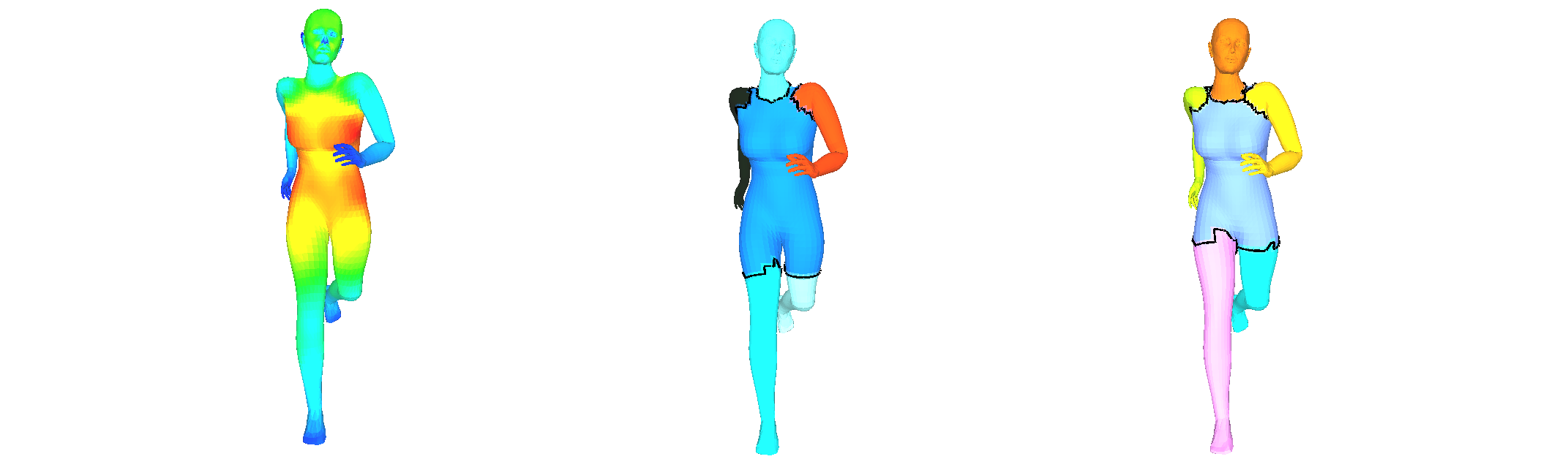}
            \caption{SDF distribution and regions comparison on woman model}
            \label{woman_comparison}
        \end{subfigure}
        \\
        \begin{subfigure}{0.5\textwidth}
            \centering
            \includegraphics[width=1\textwidth]{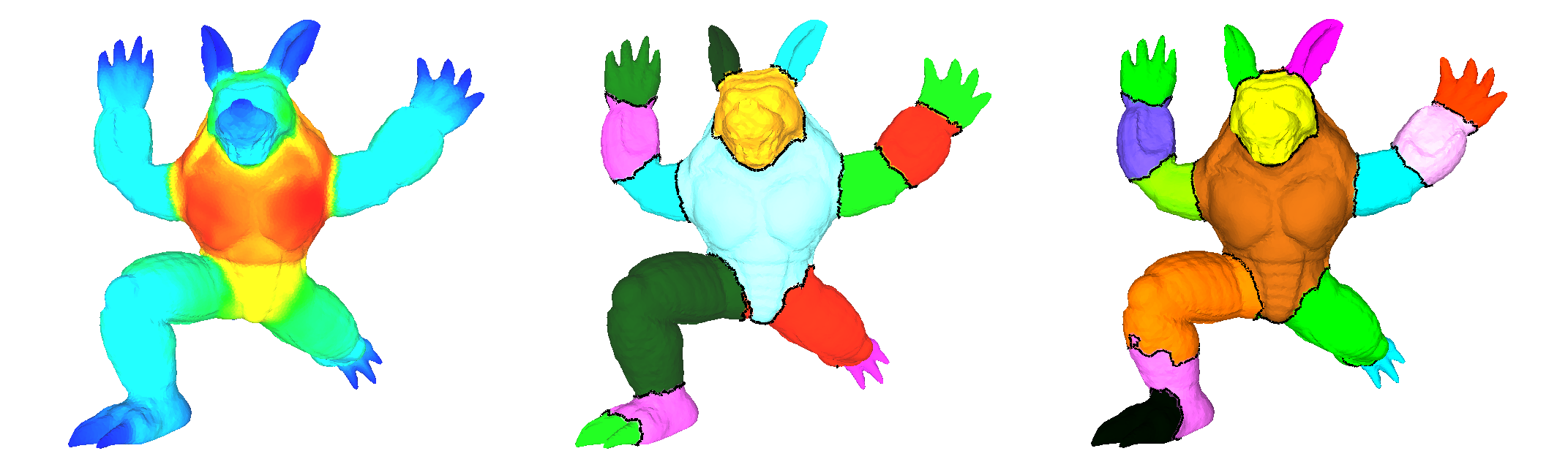}
            \caption{SDF distribution and regions comparison on armadillo model}
            \label{armadillo_comparison}
        \end{subfigure}
        \\
        \begin{subfigure}{0.5\textwidth}
            \centering
            \includegraphics[width=1\textwidth]{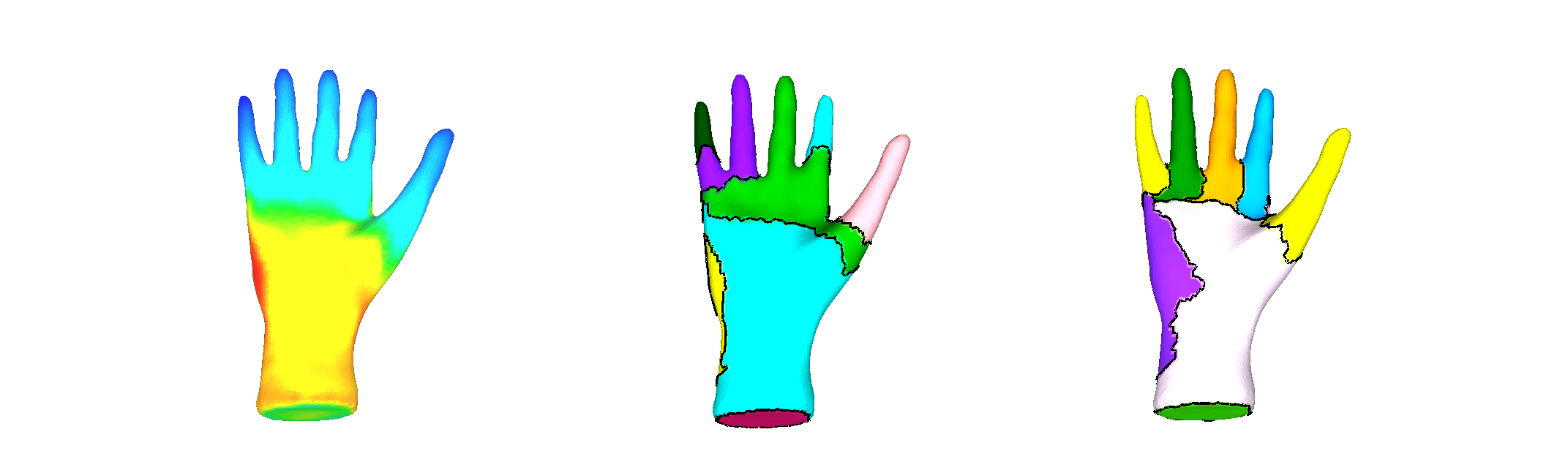}
            \caption{SDF distribution and regions comparison on hand model}
            \label{hand_comparison}
        \end{subfigure}
        \\
        \begin{subfigure}{0.5\textwidth}
            \centering
            \includegraphics[width=1\textwidth]{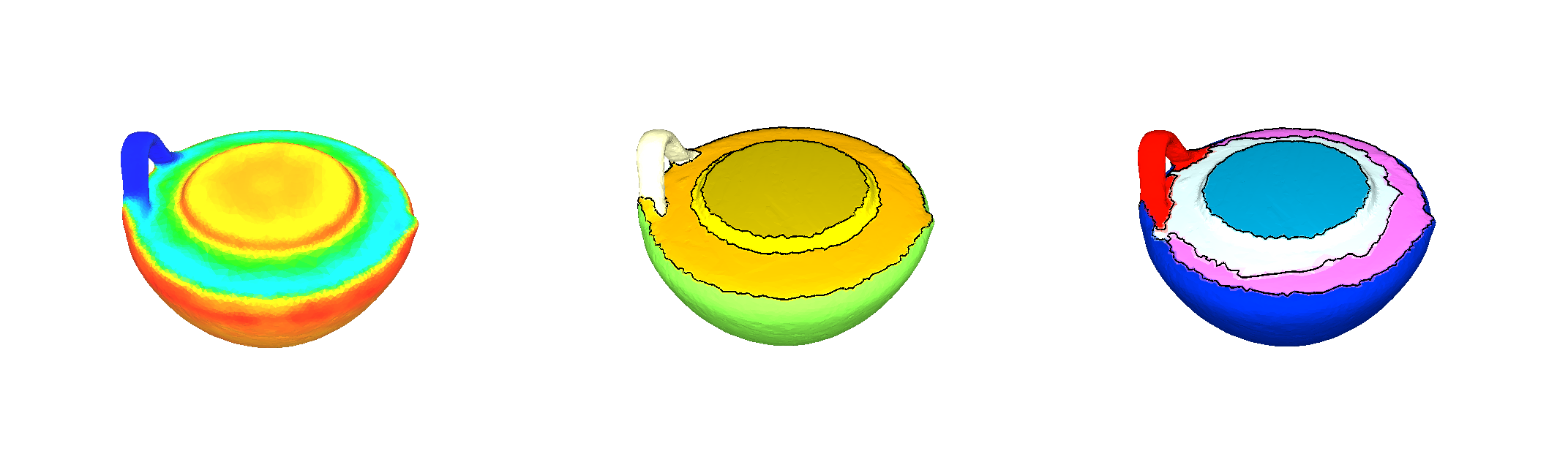}
            \caption{SDF distribution and regions comparison on pot model}
            \label{pot_comparison}
        \end{subfigure}
        \caption{Comparison of region growing costs across different models: SDF distribution (left), segmentation results with standard cost (center), and results with Reeb graph-constrained cost (right).}
        \label{growing_comparison}
    \end{figure}

    The comparative evaluation of the two proposed costs for region growing is based on four representative models of varying complexity. Figure \ref{growing_comparison} shows the mapping of SDF distribution for each model (left), followed by segmentation results with the standard cost (center) and results with the the Reeb graph-constrained cost (right).

    Figure \ref{woman_comparison} shows that the woman model presents better detection of anatomical boundaries with Reeb graph-constrained cost, particularly in the pelvic region. The armadillo model, shown in Figure \ref{armadillo_comparison}, reveals a slightly more balanced distribution of regions with cost constrained by the Reeb graph . However, the differences between costs are subtle. The differences between costs are more visible on the hand model shown in Figure \ref{hand_comparison}. The cost Reeb graph-constrained cost allows good identification of individual fingers, whereas the standard cost leads to over-segmentation of one region at the expense of the others. Finally, the last comparison, illustrated in Figure \ref{pot_comparison}, concerns the pot model. In this case, the standard cost provides a natural segmentation of the model, while the Reeb graph-constrained cost over-segments a region. The problem is that adjacent nodes on the reeb graph are not always adjacent on the mesh between saddles. The cost's double adjacency constraint can block the addition of a node to a region if it is not adjacent to it, and block the progression of a branch. The growth process then continues on the other branches, leading to over-segmentation. 

    These results highlight the importance of taking into account the topological features inherent in the Reeb graph-constrained cost when the interfaces between regions have similar scalar values. However, when the interfaces have a well-defined gradient, it is safer to use the standard cost to avoid over-segmentation due to the distribution of vertices on the mesh.
    
    \subsubsection{Results and Performance Analysis}
    
\begin{figure*}[t]
        \centering
        \includegraphics[width=1\textwidth]{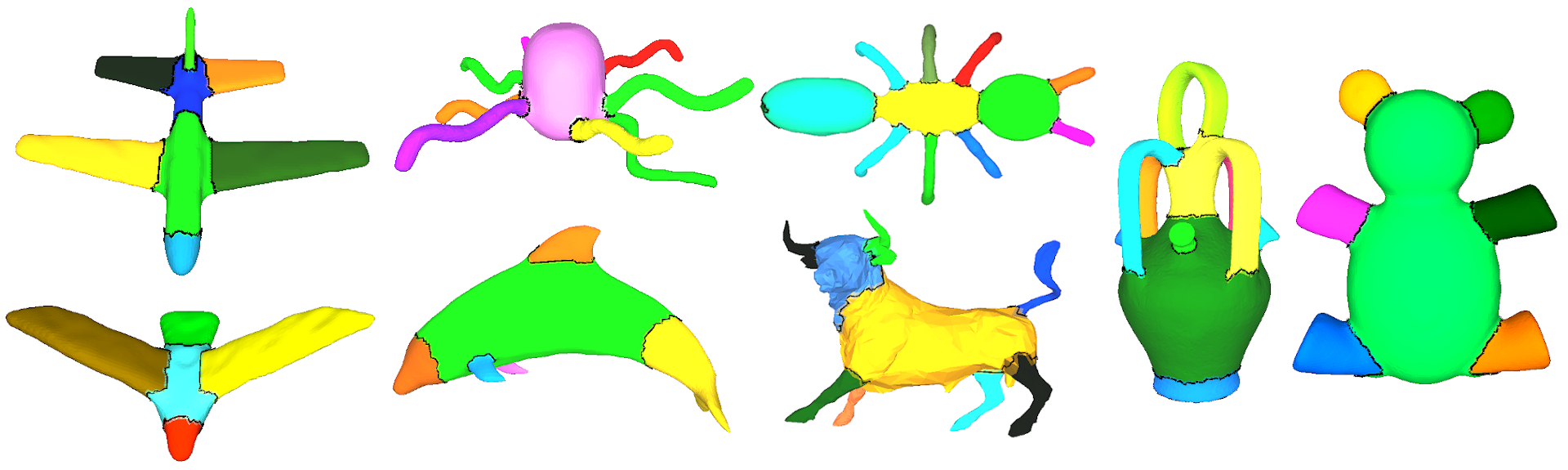}
        \caption{Segmentation results across diverse models from the Princeton Benchmark dataset using SDF. From left to right, top row: plane, octopus, ant, vase, teddy; bottom row: bird, dolphin, bull. Each color represents a distinct region identified by the method.}
        \label{fig:anthology}
    \end{figure*}

    The final evaluation concerns both the quality of the segmentation results and the computational performance of the method in this application. Figure \ref{fig:anthology} shows the results for eight representative models from the Benchmark with different shape categories. Tables \ref{tab:parameters} and \ref{tab:performance} detail the implementation parameters, number of segments and performance measures for all the models presented in this section.
    
    \begin{table}[htb]
        \caption{Implementation parameters and segmentation result}
        \label{tab:parameters}
        \centering
        \begin{tabular}{lrrrr}
            \hline
            Model & Area & Separation & Saddle Penal. & Segments \\
            \hline
            Chair & 6.0 & 10.0 & - & 12 \\
            Armadillo & 1.0 & 10.0 & 20.0 & 17 \\
            Woman & 1.0 & 15.0 & 20.0 & 6 \\
            Hand & 2.0 & 15.0 & 20.0 & 8 \\
            Pot & 2.0 & 10.0 & 20.0 & 7 \\
            Plane & 1.0 & 10.0 & - & 8 \\
            Bird & 2.0 & 10.0 & - & 5 \\
            Octopus & 2.0 & 20.0 & - & 9 \\
            Dolphin & 2.0 & 15.0 & - & 6 \\
            Ant & 1.0 & 10.0 & - & 12 \\
            Bull & 2.0 & 10.0 & - & 9 \\
            Vase & 1.0 & 10.0 & - & 6 \\
            Teddy & 2.0 & 15.0 & - & 7 \\
            \hline
        \end{tabular}
    \end{table}
    
    The segmentation results, visible in Figure \ref{fig:anthology}, indicate a semantic decomposition of the parts for most of the selected models. Standard cost has been applied to these models as they all present a clear interface between features. However, two models reveal a limitation of the method: the Teddy, which shows a head-body junction assembled in a single region instead of the expected two, and the Vase, which shows the same situation at the handle-neck transition. In both cases, intermediate SDF values in the head and neck generate saddle points rather than extrema in the Reeb graph. As a result, these zones do not form seeds for region growing phase and are absorbed by an adjacent region. These cases highlight the potential for improvement through additional geometric criteria or alternative scalar functions to achieve more semantically accurate segmentation in these regions.
    
    The parameters used to segment the models, listed in table \ref{tab:parameters}, have been adapted to the features of the shapes. SDF values were normalized by a factor of 100.  The area ratio parameter ranges from 1.0 to 6.0, with lower values preserving smaller geometrical features (e.g. 1.0 for the complex armadillo model) and higher values preserving only features above this threshold (e.g. 6.0 for the chair model). The persistence parameter, ranging from 10.0 to 20.0, controls the minimum separation of features, where lower values retain smaller topological features while higher values (e.g. 20.0 for the octopus) retain only the most pronounced features. The saddle penalty parameter indicates the penalty value applied to graph saddles with the Reeb graph-constrained cost. It indicates that the absolute value difference between two nodes must be greater than 20 to block a region in a branch and allow bypassing a saddle that has several unassigned branches. In the case of low-frequency boundaries, it allows segment boundaries to match topological features.
    
    \begin{table}[htb]
        \caption{Computational performance analysis for each steps SDF, Reeb, Cancellation and Region Growing (times in milliseconds)}
        \label{tab:performance}
        \centering
        \begin{tabular}{lrrrrr}
            \hline
            Model & Vertices & SDF & Reeb & Cancel. & Growing \\
            \hline
            Chair & 10,121 & 1,187 & 13 & 2 & 25 \\
            Armadillo & 25,273 & 7,470 & 83 & 19 & 175 \\
            Woman & 5,676 & 1,400 & 15 & 3 & 21 \\
            Hand & 6,607 & 1,392 & 16 & 2 & 26 \\
            Pot & 13,514 & 2,277 & 89 & 15 & 32 \\
            Plane & 7,351 & 764 & 12 & 2 & 24 \\
            Bird & 5,054 & 584 & 9 & 1 & 17 \\
            Octopus & 7,747 & 1,415 & 20 & 3 & 17 \\
            Dolphin & 7,121 & 1,298 & 17 & 2 & 40 \\
            Bull & 2,087 & 475 & 7 & 1 & 10 \\
            Ant & 6,376 & 785 & 9 & 1 & 16 \\
            Vase & 14,859 & 4,112 & 57 & 20 & 57 \\
            Teddy & 13,826 & 2,901 & 41 & 10 & 71 \\
            \hline
        \end{tabular}
    \end{table}
    
    Table \ref{tab:performance} shows the computation times measured at each step of the segmentation method and for all the models presented in this section. It can be seen that for all models, calculation of the scalar function dominates the processing time, ranging from 475 ms for the simplest models (Bull, 2,087 vertices) to 7,470 ms for complex geometries (Armadillo, 25,273 vertices).  Other steps are significantly faster, with Reeb graph construction completed in 7-89 ms, topological simplification in 1-20 ms and region growth in 10-175 ms. These results are consistent with the theoretical complexities of each phase given in the Methodology section. The total complexity of the segmentation method therefore depends on the scalar function used for the application.
    
	\subsection{Shape Index}
	
    \begin{figure*}[t]
         \centering
         \begin{subfigure}{0.48\textwidth}
         \centering
             \includegraphics[width=0.8\textwidth]{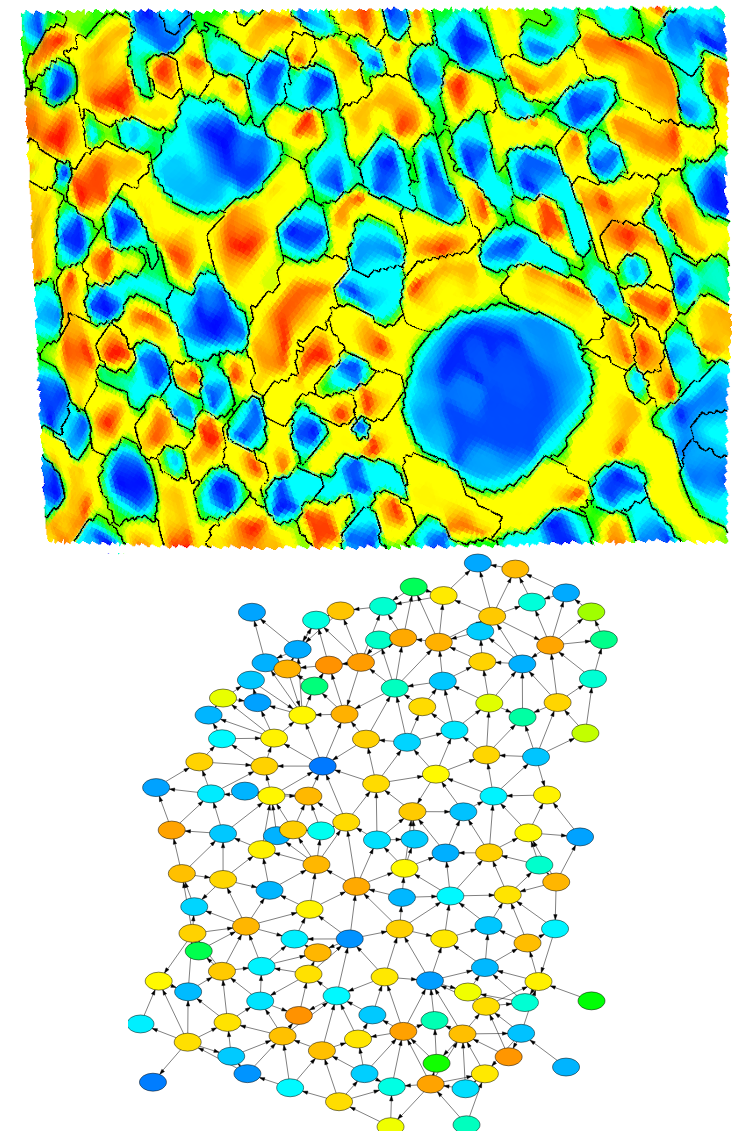}
             \caption{Area ratio 0.1 preserving fine-scale terrain features}
             \label{fig:si_detailed}
         \end{subfigure}
         \hfill
         \begin{subfigure}{0.48\textwidth}
         \centering
             \includegraphics[width=0.8\textwidth]{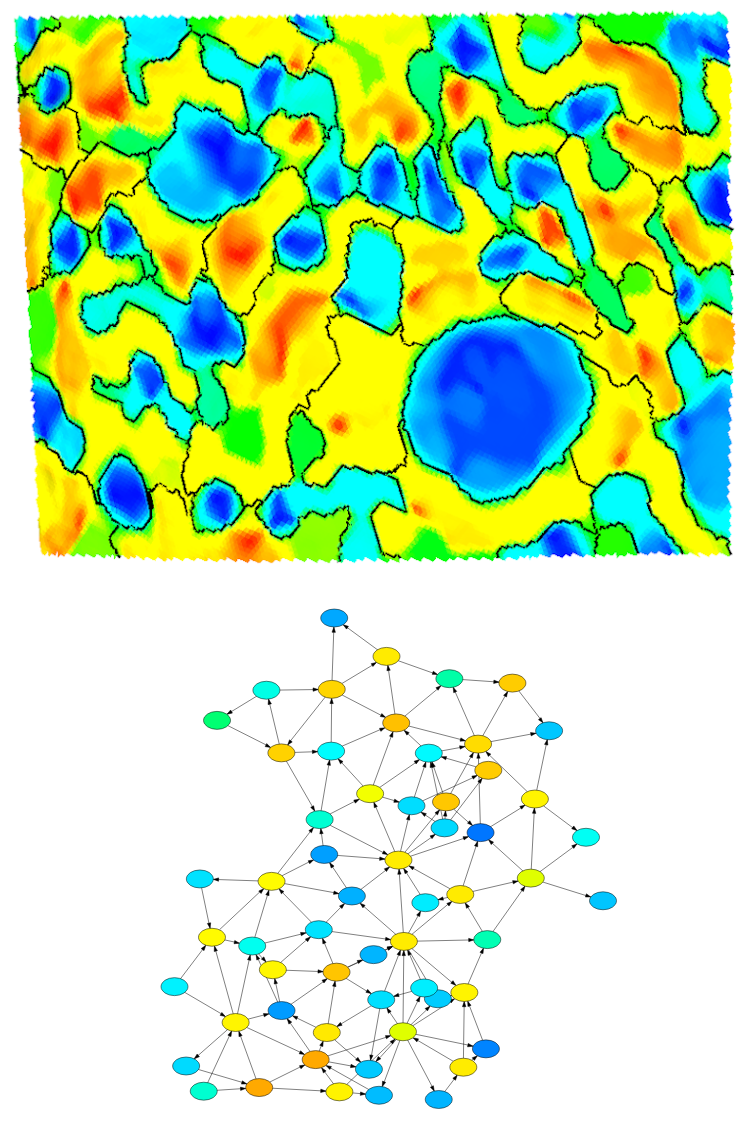}
             \caption{Area ratio 0.5 focusing on major terrain features}
             \label{fig:si_simplified}
         \end{subfigure}
            \caption{Shape Index-based segmentation of Vesta asteroid terrain showing effect of area ratio parameter on feature preservation. Top: terrain maps with segment boundaries; Bottom: corresponding adjacency graphs.}
            \label{fig:si_comparison}
    \end{figure*}
	
    The second application is a chart-based decomposition for terrain analysis using the Shape Index (SI). This is applied on a tile extracted from the asteroid Vesta \cite{Vesta}. Although height functions are traditionally used for this type of application, they have significant limitations due to their dependence on orientation and absolute altitude. This is especially true when it comes to analyzing the surface of geological bodies in space, as their curvature can mask important features and incorrectly represent local shapes.
    
    The shape index was introduced by Koenderink et al. \cite{koenderink1992surface} proposes a scale-invariant measure of local surface geometry using principal curvatures :
    \begin{equation}
    SI = \frac{2}{\pi} \arctan{\left(\frac{\kappa_2 + \kappa_1}{\kappa_2 - \kappa_1}\right)}, \quad \kappa_1 \geq \kappa_2
    \label{equ:shape_index}
    \end{equation}

    This formulation maps surface features to the interval [-1, +1], with negative values indicating concave regions and positive values representing convex areas. The normalization to [0, 1] proposed by \cite{dorai1997cosmos} facilitates computational implementation while maintaining discriminative properties. Unlike SDF which provides volumetric information, Shape Index characterizes local geometry, making it particularly valuable for planetary surface analysis where similar formations may occur at various elevations and orientations.
    
    Figure \ref{fig:si_comparison} shows the results of segmentation on a digital terrain model of the surface of asteroid Vesta (tile extracted from \cite{Vesta}). The segmented regions are delineated by black contours, with colors representing the shape index distribution of surface geometry: crater floors and depressions in blue, transitional slopes in yellow-green, and crater rims and elevated terrain in red. The segmentation uses a persistence threshold of 10 applied to the shape index (scaled by a factor of 100). The area ratio parameter varies from 0.1 in Figure \ref{fig:si_detailed} to 0.5 in Figure \ref{fig:si_simplified}, while the Reeb graph-constrained region growing uses a fixed $\lambda$ value of 20. The corresponding region adjacency graphs are displayed below each model.
    
    The Reeb graph-constrained cost is needed for chart-based applications of this type, as the slope transitions between features share similar value ranges, making the standard cost ineffective for differentiation. The method integrates both geometrical and topological features of the shape index to produce region contours that naturally align with the shape's morphology, effectively separating convex and concave regions.
    
    Varying the area ratio parameter shows the smoothing of scalar values from small features, which merge into regions of larger features while preserving the values of the latter. This property indicates that the topological simplification algorithm can be used as a smoothing approach for scalar functions, removing undesired features while preserving significant ones. Furthermore, the boundaries of persistent regions remain relatively stable, as the Reeb graph-constrained cost use topological features to propagate regions. The adjacency graph retains its structural distribution despite containing fewer nodes, forming a simplified mesh representation. This hierarchical organization can enable multi-scale analysis through area ratio parameter adjustment. 

\section{Conclusion}

    This paper presents a flexible mesh segmentation method that integrates geometrical and topological features through Reeb graph representation. The main contributions of the method include a new critical node cancellation algorithm that preserves vertex-graph correspondence while handling both morse and degenerate nodes, and a region growing algorithm that guarantees region continuity via two different cost functions. The method is also optimized to achieve an overall complexity of O(n log n). It is validated through two different applications: part-based decomposition using the Shape Diameter function and chart-based decomposition for terrain analysis using Shape Index.
    
    Evaluation of these applications shows both the flexibility and the limitations of the method. For part-based segmentation using SDF, the method successfully decomposes most models into meaningful components, particularly for shapes with well-defined geometric transitions. However, it shows limitations in regions with intermediate SDF values at feature transitions, as shown by the merged regions in the Teddy and Vase models. This is because these intermediate values generate saddle points rather than extrema in the Reeb graph, preventing the creation of new region seeds. In chart-based decomposition using Shape Index, the method effectively segments terrain features at different scales. The area ratio parameter controls the granularity of feature detection, enabling analysis of fine-scale terrain details down to major topographic features. Comparison of growth costs indicates that the standard cost is optimal when features have clear geometric boundaries, while the cost constrained by the Reeb graph, allows better definition of boundaries when transitions are gradual, as in chart-based applications. However, the latter can lead to over-segmentation when mesh and graph adjacencies differ significantly, underlining the importance of cost selection according to application context.
    
    Future work could extend the approach to multivariate Reeb graphs to enable simultaneous analysis of complementary scalar functions when a single function fails to capture all relevant features. The development of automatic parameter estimation methods based on mesh characteristics would also improve the accessibility of the method while maintaining its flexibility. Finally, the use of scalar functions that support chart-based decomposition, such as the shape index, opens new possibilities for mesh simplification and comparative shape analysis using region adjacency graphs.

\section*{Acknowledgements}
This project has been funded by the Burgundy Regional Council under the contract 2022-Y-14262 "3DKAR".

\bibliographystyle{eg-alpha-doi} 
\bibliography{egbibsample}       


\end{document}